\pdfoutput=1
\documentclass[11pt]{article}

\usepackage[margin=1in]{geometry}
\usepackage{amsmath,amssymb,amsthm}
\usepackage{graphicx}
\usepackage{booktabs}
\usepackage{adjustbox}
\usepackage[round,authoryear]{natbib}
\usepackage[hidelinks]{hyperref}

\graphicspath{{figures/}}

\newcommand{\E}{\mathbb{E}}
\newcommand{\KL}{\mathrm{KL}}
\newcommand{\F}{\mathcal{F}}
\newcommand{\G}{\mathcal{G}}

\newtheorem{theorem}{Theorem}
\newtheorem{proposition}{Proposition}
\newtheorem{corollary}{Corollary}
\theoremstyle{remark}
\newtheorem{remark}{Remark}

\title{Betting against the simulator:\\
Monte Carlo e-processes under adaptive actions}

\author{Elena Bortolato\\ CUNEF Universidad}

\date{  \today}

\begin{document}
\maketitle

\begin{abstract}
Multi-arm multi-stage trials, bandits and digital twins collect data
adaptively: which arm is observed depends on what has been seen. We test whether a simulator reproduces such a
system with e-processes, which allow optional stopping. A score learned by classifying real against simulated
outcomes becomes an exact bet when normalised over $M$ fresh simulations. Under adaptive allocation, validity follows from
exchangeability if the simulator is run under the action taken and from the
current state. This conditioning is also necessary: a simulator that ignores
the action makes the expected bet tend to $1+\chi^2$, and the wealth grows
exponentially under the null. The actions are moreover a source of power: in
an audit of a language-model API, choosing the prompts
adaptively makes every score faster by
up to 61\% in the number of queries. The growth rate of any score splits into
a simulation cost, exactly $M/(M+1)$ for local alternatives, and a learning
cost, $\KL(Q\|\tilde T)$, where $\tilde T$ is the score's forecast.
Treated as a likelihood in the score's parameter, the bet admits exact
Bayesian updating: posterior-predictive bets stay within $d\log t$ of the best score in hindsight on
every sample path. In a multi-arm multi-stage trial of insulin-dosing strategies with
re-assignment of participants, the monitor controls
false alarms and identifies the arms on which the simulator fails; monitors
that ignore the allocation or the participants' history do not.
\end{abstract}

\section{Introduction}\label{sec:intro}

A digital twin is useful only as long as it behaves like the system it
copies. Checking this is hard when the system is being steered while we
watch. Consider a multi-arm multi-stage (MAMS) trial of insulin-dosing
strategies \citep{whitehead1997design,magirr2012generalized}. At each interim
analysis some arms are dropped and their participants move to other arms,
taking their glucose history with them. A simulator of each participant's
glucose dynamics is available, and the question is whether, and when, the
real participants stop behaving like simulated ones.

The same question comes up with multi-armed bandits
\citep{lattimore2020bandit}, response-adaptive clinical designs
\citep{trippa2012bayesian,villar2015multi}, and AI systems that act in
environments shaped by their own past decisions
\citep{perdomo2020performative,dulacarnold2021challenges}. For example, a
content-ranking or moderation system changes what users post and what the
platform observes next, so the data distribution depends on the policy
deployed so far \citep{chaney2018algorithmic,jiang2019degenerate}; such
systems are routinely designed and tuned on simulators of user behaviour
\citep{ie2019recsim,shi2019virtual}, whose fidelity to the real platform is
exactly what is in doubt \citep{zhao2020sim}. In these settings, as we choose
actions based on past observations, can we continuously test whether the
resulting stream of outcomes is still consistent with a simulator, and sound
an alarm as soon as it is not, while controlling false alarms under optional
stopping? The data are dependent, because what is observed next depends on
what has been seen; the null is a simulator whose likelihood we
cannot write down; and we want to look at the evidence after
every observation. Fixed-sample tests lose
their guarantees under repeated looks, and the corrections developed for
adaptive trials \citep{bowden2017unbiased,magirr2012generalized} assume a
pre-specified adaptation rule. Bayesian updating copes with adaptive sampling
but does not, by itself, control the type~I error under optional stopping.

E-values suit this situation
\citep{vovk2021evalues,ramdas2023game,grunwald2024safe}. An e-process is the
running wealth of a gambler who places one fair bet against the null per
observation. By Ville's inequality, the probability that this wealth ever
exceeds $1/\alpha$ is at most $\alpha$, so the experimenter can look after
every observation and stop at any time. The best known e-process is the Bayes
factor. Written one observation at a time, it is a product of ratios: the
posterior predictive density of each new observation over its null density.
Each factor has mean one under the null for every prior, and the product
rewards whoever forecasts the data better, the learner or the null
\citep{dawid1984present}. Learning the parameters along the way costs only a
term of order $\log t$ \citep{clarke1990information}.

We cannot use the Bayes factor here, for two reasons. The null density is not
available, since the simulator has latent dynamics and, under adaptive
allocation, its law changes with the path. Nor is there a trusted model of the
real system to forecast with, since that is precisely what is in question.
Our approach keeps the shape of the Bayes factor, one fair bet per
observation placed before the observation is seen, and replaces both of its
ingredients.

To replace the null density, we start from a classifier. Trained to
distinguish real outcomes from simulated ones, a classifier estimates the
ratio of their densities from samples alone \citep{thomas2022lfire}. Call
\emph{score} any positive function $T$ of an outcome that is large where real
outcomes are more likely than simulated ones; the classifier's estimate of the
density ratio is the natural choice. A score is not a fair bet by itself, but
a Monte Carlo test turns it into one \citep{besag1989generalized}. At each
step $t$ we observe the real outcome $x_t$, run the simulator $M$ times to
obtain $y_{t,1},\dots,y_{t,M}$, and bet $M+1$ times the real outcome's share
of the total score. If the simulator is right, the real outcome is just one
more draw among $M+1$ exchangeable ones, and the bet is fair for any score. In
an adaptive design the only delicate point is how to simulate. We run the
simulator under the action actually taken and from the current state of the
unit. This is enough for validity under any allocation rule, and it cannot be
dropped: a simulator run without regard to the action produces false alarms
with probability one.

The score also replaces the forecast. Multiplying the simulator's density by
the score and normalising gives a density for the next outcome, which moves
the simulator's mass towards where the score expects real outcomes to fall:
choosing a score amounts to forecasting reality. The score is unknown and must
be learned as the data arrive, so we let it depend on parameters $\theta$, for
example $T_\theta(x) = \exp\{\theta^\top z(x)\}$ for a vector of features
$z(x)$. The bet is itself a likelihood for $\theta$. Up to the factor $M+1$,
the bet is the real outcome's share of the total score,
$T_\theta(x_t)/\{T_\theta(x_t) + \sum_{j} T_\theta(y_{t,j})\}$: the
probability that a conditional-logit model \citep{mcfadden1974conditional},
which picks one item from a set
with probability proportional to its score, picks the real outcome among the
$M+1$ candidates. The product of these probabilities over time is a likelihood
for $\theta$, so $\theta$ can be learned by Bayes' rule: we put a prior on
$\theta$, update the posterior after each observation, and place the next bet
with the posterior average of the bet. Since the posterior uses only past
observations, this average is still a fair bet under the null. It plays the
role of the posterior predictive in the Bayes factor, with the score in place
of the model. Standard regret bounds for Bayesian mixtures
\citep{kakade2005online} then guarantee that, on every sample path, the wealth
stays within a $d\log t$ term of what the best score in the class would have
earned in hindsight.

Our contributions are the following.
\begin{enumerate}
\item \emph{Validity under adaptive actions, and why the conditioning is
needed.} The Monte Carlo e-process is a test martingale for every allocation
rule (Theorem~\ref{thm:valid}). Without conditioning on the action, the
expected bet under the null tends to $1+\chi^2$ for the likelihood-ratio score
and false alarms grow exponentially, at a rate of at most $\log K$ per step
with $K$ arms (Proposition~\ref{prop:necessary},
Corollary~\ref{cor:falsealarm}).
\item \emph{Bayesian updating of the score.} Posterior-predictive bets,
updated in one step per observation, with a regret bound on every sample path
(Theorem~\ref{thm:simlik}); approximate posteriors keep validity exact at a
bounded cost in power (Proposition~\ref{prop:approx}).
\item \emph{Adaptive actions as a source of power.} Since validity holds for
any allocation rule, the actions can be chosen to collect evidence quickly. In
an audit of a language-model API, choosing the prompts adaptively reduces the
number of queries by up to 61\%, for our scores and for those of concurrent
methods (Section~\ref{sec:exp-llm}).
\item \emph{The power of an arbitrary score.} A score implicitly bets that
reality looks like the simulator reweighted by the score, a tilted version of
the null. Its growth rate is the Kullback--Leibler information against the
simulator minus the distance from reality to this tilted null, and $M$
simulations keep a fraction $M/(M+1)$ of it for local alternatives
(Theorem~\ref{thm:growth}).
\end{enumerate}

Validity and power are separate questions in this construction. Validity
comes from how the simulator is run, not from the score, so any score fixed
before the new outcome is seen gives a valid test, and a network trained in
advance can be used in the same adaptive design. Our update rules are what we
recommend when nothing is known about the alternative, which is the usual
situation when a simulator is validated: a twin tends to fail in ways its
designers did not anticipate, and a score tuned to a guessed departure has no
power against another.

In the insulin trial, with five strategies over three stages, only the monitor
that runs the twin from each participant's state under the assigned arm
controls false alarms (2\% of trials, against 95\% for a twin that ignores the
participants' history). One e-process per arm, combined by e-BH
\citep{xu2021unified}, identifies the arm on which the twin fails.

Three recent papers use the same Monte Carlo bet for independent data
\citep{dombowsky2026besag,forre2026anytime,cortinovis2026slice}; for a single
test it is the soft-rank e-value of \citet{wang2022false}. We developed our
construction independently (Appendix~\ref{app:timeline}). Their main
developments assume independent observations and a simulator without state.
\citet{forre2026anytime} list dependent data as an open problem, and
\citet[Appendix~C.1.1]{cortinovis2026slice} note that the construction extends
under conditional exchangeability and prove validity given independent
covariates. They also learn the score differently: \citet{forre2026anytime}
refit it by maximising its empirical growth, and \citet{cortinovis2026slice}
train a neural ratio estimator before the data arrive, over the simulator's own
parametric family, so that their score can forecast only departures that the
family contains; our scores reweight the simulator by features of the outcome,
and detect departures that its designers did not build into it.
Section~\ref{sec:discussion} lists in detail which of our results coincide
with theirs.

Section~\ref{sec:method} gives the method and its validity,
Section~\ref{sec:power} its power, and Section~\ref{sec:predictive} the
predictive view and the update rules (Algorithm~1). Section~\ref{sec:experiments}
contains simulations, the comparison with concurrent methods and the
language-model audit, and Section~\ref{sec:trial} the trial.
Section~\ref{sec:discussion} discusses limitations.

\section{Monte Carlo e-processes under adaptive control}\label{sec:method}

\subsection{Setting}\label{sec:setting}

The symbols used throughout are collected in Table~\ref{tab:notation} in the
Appendix.

At each time $t = 1, 2, \ldots$ an action (or context) $A_t$ is chosen,
possibly at random, by a policy that may depend on everything observed
before, including the monitor's own wealth. The real system then produces an
outcome $X_t$ with conditional law $Q_t(\cdot \mid \F_{t-1}, A_t)$, and the
analyst can draw samples from a simulator with conditional law
$P_t(\cdot \mid \F_{t-1}, A_t)$; for a digital twin this means re-initialising
the twin at the current state and running it under action $A_t$. The
filtration $\F_t$ contains the actions, the real outcomes, the simulator draws
and any internal randomisation up to time $t$, and
$\G_t = \F_{t-1} \vee \sigma(A_t)$ is the information available when the bet
at time $t$ is placed. The null hypothesis is
\[
H_0:\quad Q_t(\cdot \mid \F_{t-1}, A_t) = P_t(\cdot \mid \F_{t-1}, A_t)
\quad \text{for all } t, \text{ almost surely}:
\]
the simulator reproduces the conditional behaviour of the system under every
action that is actually taken. $H_0$ is composite in the policy, which is left
unrestricted. Because each factor is fair conditionally on the past and on
the action, one e-process is valid for every policy at once, with no infimum
over the class. The policy may depend
on past outcomes and on the monitor's own wealth; the price is that
the null concerns only the actions actually taken, so the monitor cannot
detect failures of the simulator under actions the policy never chooses. A nonnegative process $(W_t)$ with $W_0 = 1$ is a \emph{test
supermartingale} for $H_0$ if $\E[W_t \mid \F_{t-1}] \le W_{t-1}$ under every
distribution in $H_0$; by Ville's inequality, rejecting the first time $W_t$
reaches $1/\alpha$ controls the type-I error at level $\alpha$ uniformly over
time and stopping rules \citep{ramdas2023game,grunwald2024safe}. Products of
nonnegative factors with
\begin{equation}\label{eq:fair}
\E[E_t \mid \G_t] \le 1 \quad \text{under } H_0
\end{equation}
are test supermartingales.

\subsection{The Monte Carlo e-process}

Let $s_t$ be a real-valued score chosen from the information in $\G_t$, and
$T_t = e^{s_t}$. Draw $Y_{t,1}, \ldots, Y_{t,M}$ from the simulator under
action $A_t$ and set
\begin{equation}\label{eq:mc}
E_t = \frac{(M+1)\, T_t(X_t)}{T_t(X_t) + \sum_{j=1}^{M} T_t(Y_{t,j})}.
\end{equation}

\begin{proposition}[One-step exactness; \citealp{wang2022false}; \citealp{dombowsky2026besag}, Proposition~1]\label{prop:exact}
Let $U_0, \ldots, U_M$ be exchangeable and $T > 0$. Then
$\E\big[(M+1)\,T(U_0)/\sum_{k} T(U_k)\big] = 1$.
\end{proposition}

\begin{proof}
The ratios $T(U_k)/\sum_{k'} T(U_{k'})$ sum to one and, by exchangeability,
have equal expectations, each equal to $1/(M+1)$.
\end{proof}

The normalised scores $T_t(U_k)/\sum_{j} T_t(U_j)$ can be viewed as the
probabilities that the
score assigns to each of the $M+1$ outcomes being the real one, and $E_t$ is
$M+1$ times the probability it assigns to the outcome that is actually real.
The bet pays when the score picks out the real outcome better than a uniform
guess, $1/(M+1)$; under the null the real outcome is indistinguishable from
the simulations, so on average no score can.

A score is typically a classifier's estimate of the ratio $q_t/p_t$ of the
densities of the real outcome and of the simulator (Section~\ref{sec:scores}).
Used directly as the bet, $T_t(X_t)$ (the \emph{plug-in} bet) has conditional
mean $Z_t = \E_{P_t}[T_t]$ under the null, which equals one only if the
estimated ratio integrates exactly to one against the simulator. A learned
ratio does not. If $\log T_t = \log(q_t/p_t) + \varepsilon$ with
$\E[\varepsilon] = 0$, then $\E[Z_t] = \E[e^{\varepsilon}] > 1$ by Jensen's
inequality even when $q_t = p_t$, so estimation noise alone makes the plug-in
bets favourable under the null; this is why an overfitted learner gives many
false alarms (Section~\ref{sec:exp-drift}). The factor \eqref{eq:mc} replaces
$Z_t$ by its estimate
$\widehat Z_t = \{T_t(X_t) + \sum_{j=1}^{M} T_t(Y_{t,j})\}/(M+1)$, and
Proposition~\ref{prop:exact} shows that this self-normalisation makes the
conditional mean exactly one, whatever $Z_t$ is.

The factor is bounded, $0 < E_t \le M+1$. To hedge the fraction of wealth that
is bet, we average the processes $\prod_i (1 - \lambda + \lambda E_i)$ over a
finite grid $\Lambda \subset (0, 1]$,
\begin{equation}\label{eq:wealth}
W_t = \frac{1}{|\Lambda|} \sum_{\lambda \in \Lambda} \prod_{i \le t}
\big(1 - \lambda + \lambda E_i\big).
\end{equation}

\begin{theorem}[Anytime validity under adaptive control]\label{thm:valid}
Suppose that, conditionally on $\G_t$, the simulator draws
$Y_{t,1}, \ldots, Y_{t,M}$ are i.i.d.\ from $P_t(\cdot \mid \F_{t-1}, A_t)$ and
independent of $X_t$, and that the score $s_t$ is $\G_t$-measurable. Then
under $H_0$, $\E[E_t \mid \G_t] = 1$ for every $t$, the wealth
\eqref{eq:wealth} is a test martingale, and
$P(\exists t: W_t \ge 1/\alpha) \le \alpha$. This holds for every policy,
every learner and every $M \ge 1$.
\end{theorem}

\begin{proof}
Under $H_0$ and conditionally on $\G_t$, the real outcome $X_t$ has law
$P_t(\cdot \mid \F_{t-1}, A_t)$ and is independent of the simulator draws, so
$(X_t, Y_{t,1}, \ldots, Y_{t,M})$ is i.i.d.\ and hence exchangeable, while
$T_t$ is fixed. Proposition~\ref{prop:exact} gives
$\E[E_t \mid \G_t] = 1$, so $\E[1 - \lambda + \lambda E_t \mid \F_{t-1}] = 1$
for each $\lambda$. Each product is a nonnegative martingale with initial
value one, and so is their average. Ville's inequality concludes.
\end{proof}

For an i.i.d.\ null without actions, Theorem~\ref{thm:valid} reduces to the
Besag--Clifford e-process of \citet[Proposition~6]{dombowsky2026besag}. The
theorem involves no density, no property of the score and no restriction on
the policy, which may depend on the data and on the wealth itself; in a MAMS
trial the action $A_t$ is the arm assigned to a participant for a period,
chosen after interim analyses and re-assignments, and the conditional law
$P_t$ is that of the participant's outcome given their history. The policy
enters only through the requirement that the simulator be run under the action
actually taken. Proposition~\ref{prop:necessary} shows that this
requirement cannot be dropped.

\begin{proposition}[Conditioning on the action cannot be dropped]\label{prop:necessary}
Suppose that $H_0$ holds and that at time $t$ the policy takes an action $a$,
with $P_a = P_t(\cdot \mid \F_{t-1}, a)$, while, conditionally on $\G_t$, the
simulator draws $Y_{t,1}, \ldots, Y_{t,M}$ are i.i.d.\ from a law $\bar P$
that does not depend on $A_t$ (for instance the twin run under uniformly
random actions) and are independent of $X_t$. Let $T > 0$ be a bounded
$\G_t$-measurable score.
\begin{enumerate}
\item[(i)] \emph{The limit of the expected bet.} As $M \to \infty$,
\[
\E[E_t \mid \G_t] \to \rho = \frac{\E_{P_a}[T]}{\E_{\bar P}[T]}
\quad\text{and}\quad
\E[E_t^2 \mid \G_t] \to \frac{\E_{P_a}[T^2]}{(\E_{\bar P}[T])^2}.
\]
\item[(ii)] \emph{Favourable scores exist whenever $P_a \ne \bar P$.} The
limit $\rho$ equals one for every bounded score if and only if $P_a = \bar P$.
If $P_a \ll \bar P$ and $R_a = dP_a/d\bar P$, the likelihood-ratio score
gives
\[
\E_{P_a}[R_a] = \int R_a^2\, d\bar P = 1 + \chi^2(P_a \,\|\, \bar P):
\]
$T = R_a$ has $\rho = 1 + \chi^2(P_a\|\bar P)$ when $R_a$ is bounded and
bounded away from zero, and in general the truncated scores
$T_c = \min(R_a, c) + 1/c$ have $\rho \to 1 + \chi^2(P_a\|\bar P) \in (1,\infty]$
as $c \to \infty$.
\item[(iii)] \emph{Hence false alarms.} Suppose that $P_a$ and $\bar P$ do not
change over time and that, on an event $\Omega_0$, from some finite time on
the policy plays $a$ with a fixed score $T$ taking values in $[b, B]$,
$0 < b \le B$, whose limit in (i) is $\rho > 1$; by (ii) such a score exists
whenever $P_a \ne \bar P$. Then there is $M_0$ such that, for every
$M \ge M_0$, if $\Lambda$ contains some
$\lambda \le \min\{1/2,\, (\rho - 1)b^2/(8B^2)\}$, on $\Omega_0$, almost
surely,
\[
\liminf_{t\to\infty} \frac{\log W_t}{t} \ge \frac{\lambda(\rho - 1)}{4} > 0 .
\]
In particular the test rejects the true null with probability at least
$P(\Omega_0)$.
\end{enumerate}
\end{proposition}

The proof is in Appendix~\ref{app:proofs}. The event $\Omega_0$ occurs, for
example, when a bandit settles on the best arm: each bet then has mean larger
than one, by (i) and (ii), and the wealth grows exponentially, by (iii). The
proposition also shows that the condition $P_a = \bar P$ is not only
sufficient for fairness (Theorem~\ref{thm:valid}) but necessary for the bet to
be fair against every score, and that $1 + \chi^2(P_a\|\bar P)$ is the exact
limit of the expected bet of the likelihood-ratio score.

For a single Monte Carlo test, simulations drawn from a law within total
variation $\varepsilon$ of the right one inflate the level by at most
$\varepsilon$ \citep[Theorem~2]{barber2026montecarlo}. In a sequential
e-process the bias of every factor compounds instead: the expected bet is
governed by the $\chi^2$ divergence and the growth of the wealth by the
Kullback--Leibler divergence (Corollary~\ref{cor:falsealarm}), and
$2\,\mathrm{TV}^2 \le \KL \le \log(1+\chi^2)$, so any departure, however small in
total variation, eventually raises a false alarm. Moreover $\chi^2$ can be
large when the total variation is small, if the real system produces rare
outcomes that the simulator almost never does.

\section{Power: the growth rate}\label{sec:power}

The time needed to reach $1/\alpha$ is governed
by the growth rate $\E[\log E_t \mid \G_t]$ \citep{grunwald2024safe}. Fix a
time, condition on $\G_t$ and drop the index. Let $P$ be the simulator
law and $Q$ the true law of the next outcome, with $Q \ll P$; let $T = e^{s}$
and let $Z = \E_P[T] < \infty$ be the normalising constant of
Section~\ref{sec:method}. The score reweights the simulator law into a new
law for the next outcome, its \emph{tilted null},
\begin{equation}\label{eq:tilt}
\frac{d\tilde T}{dP} = \frac{T}{Z},
\end{equation}
which keeps the outcomes of the simulator and gives each a weight
proportional to its score: a change of measure on the space of outcomes, not
the distribution of the score values. It has density $Tp/Z$ when $P$ has
density $p$, and we use the same symbol for the law and its density. It is the
score's forecast of $Q$, and a perfect score has $\tilde T = Q$
(Section~\ref{sec:scores} develops this reading). Write
$\rho = \E_Q[T]/Z$, the limit of the expected bet as $M \to \infty$
(Proposition~\ref{prop:necessary}(i), with $Q$ and $P$ in place of $P_a$
and $\bar P$). Let $X \sim Q$
and, independently, $Y_1, \ldots, Y_M$ i.i.d.\ $P$, and let
$E^{(M)} = (M+1)T(X)/\{T(X) + \sum_{j} T(Y_j)\}$ be the factor \eqref{eq:mc}
computed from them. The \emph{growth rate} with $M$ simulator draws is
\begin{equation}\label{eq:growthrate}
\begin{aligned}
g_M = g_M(Q, P, T) &= \E_{Q \otimes P^M}\big[\log E^{(M)}\big] \\
&= \int \log\frac{(M+1)\,T(x)}{T(x) + \sum_{j=1}^M T(y_j)}\;
Q(dx) \prod_{j=1}^M P(dy_j),
\end{aligned}
\end{equation}
the expected logarithm of one factor when the whole wealth is bet. Since
$\log E^{(M)} \le \log(M+1)$ it is well defined in $[-\infty, \log(M+1)]$, and
finite when $s$ is bounded. It is at most zero when $Q = P$, by Jensen's
inequality and Proposition~\ref{prop:exact}. When a fraction $\lambda$ of the
wealth is bet, as in each term of \eqref{eq:wealth}, the rate is
\[
g_M(\lambda) = g_M(\lambda; Q, P, T)
= \E_{Q \otimes P^M}\big[\log(1 - \lambda + \lambda E^{(M)})\big],
\qquad g_M = g_M(1).
\]
In the sequential procedure $Q$, $P$ and $T$ are the conditional laws and the
score at time $t$, and the \emph{conditional growth rate}
$g_{M,t} = \E[\log E_t \mid \G_t]$ is $g_M$ evaluated at them; it is a
random variable that depends on the history. We write $g_M$ without arguments
when $Q$, $P$ and $T$ are clear from the context.

\begin{theorem}[Growth rate]\label{thm:growth}
Assume $\KL(Q\|P) < \infty$ and $\rho < \infty$.
\begin{enumerate}
  \item[(i)] (Ceiling.) $g_M \le \KL(Q\|P)$ for every $M$ and every score.
  \item[(ii)] (Unlimited budget.) As $M \to \infty$,
        $E^{(M)} \to (d\tilde T/dP)(X)$ almost surely, and if $s$ is bounded,
        $g_M \to g_\infty = \E_Q[\log (d\tilde T/dP)(X)] = \KL(Q\|P) - \KL(Q\|\tilde T)$.
  \item[(iii)] (Finite budget.) $g_M \ge g_\infty - \log\{1 + (\rho - 1)/(M+1)\}
        \ge g_\infty - (\rho - 1)/(M+1)$.
  \item[(iv)] (Local alternatives.) If $dQ/dP = 1 + \varepsilon \psi$ and
        $T = e^{\varepsilon h}$ with $\psi, h$ bounded and $\E_P[\psi] = 0$, and
        $\tilde h = h - \E_P[h]$, then
        $g_M = \frac{M}{M+1} \cdot \frac{\varepsilon^2}{2}
        \big(\E_P[\psi^2] - \E_P[(\psi - \tilde h)^2]\big) + O(\varepsilon^3)$,
        while $\KL(Q\|P) = \frac{\varepsilon^2}{2}\E_P[\psi^2] + O(\varepsilon^3)$.
\end{enumerate}
\end{theorem}

Part (i) is the log-optimality of the likelihood ratio. When the score is the
likelihood ratio, $\tilde T = Q$ and (ii) is the limit in Theorem~1 of
\citet{dombowsky2026besag}; part (iii) is their Theorem~3, rewritten with the
tilted null. What is new is the reading of an arbitrary score through
$\tilde T$ in (ii), and the local constant in (iv), which factorises the
efficiency relative to the oracle likelihood ratio into a simulation term and
a learning term,
\[
\frac{g_M}{\KL(Q\|P)} \approx
\underbrace{\frac{M}{M+1}}_{\text{simulation}} \times
\underbrace{\Big(1 - \frac{\KL(Q\|\tilde T)}{\KL(Q\|P)}\Big)}_{\text{learning}}.
\]
One draw per observation halves the growth rate and so roughly doubles the time
to detection; nine draws keep 90\% and sixteen 94\%. Away from the local regime,
$M/(M+1)$ is neither an upper nor a lower bound: for a large mean shift and a
perfect score the efficiency at $M = 1$ is lower, while an overconfident score
does better with a finite budget than with an infinite one, because
$E^{(M)} \le M+1$ truncates its bets (Section~\ref{sec:exp-budget}).

Theorem~\ref{thm:growth} describes one factor. Over the whole sequence, the
log-wealth is at least the sum of the conditional growth rates of its factors,
up to fluctuations that average out, so the test rejects once that sum exceeds
$\log(1/\alpha)$.

\begin{corollary}[Cumulative growth]\label{cor:cumulative}
For $\lambda \in \Lambda$ let
$g_{M,i}(\lambda) = \E[\log(1 - \lambda + \lambda E_i) \mid \G_i]$ be the
conditional growth rate of the $\lambda$-term of \eqref{eq:wealth} at time $i$.
\begin{enumerate}
\item[(i)] For every $\lambda \in \Lambda$ with $\lambda < 1$, and with no
condition on the scores,
$\log W_t \ge \sum_{i \le t} \log(1 - \lambda + \lambda E_i) - \log|\Lambda|$,
each increment lies in $[\log(1-\lambda),\, \log(1 + \lambda M)]$, and almost
surely
\[
\liminf_{t \to \infty} \frac1t \Big(\log W_t - \sum_{i \le t} g_{M,i}(\lambda)\Big) \ge 0 .
\]
\item[(ii)] If $1 \in \Lambda$ and the scores are bounded, $|s_t| \le B$, the
same holds for $\lambda = 1$, with increments $\log E_i \in [-2B, \log(M+1)]$
and rates $g_{M,i} = g_{M,i}(1)$.
\end{enumerate}
If, for some $\lambda \in \Lambda$, the average of $g_{M,i}(\lambda)$ over
$i \le t$ stays above $\bar g > 0$, the test rejects with probability one,
after about $(\log(1/\alpha) + \log|\Lambda|)/\bar g$ steps.
\end{corollary}

\begin{proof}
The average \eqref{eq:wealth} is at least $|\Lambda|^{-1}$ times any one of its
terms. Since $0 < E_i \le M+1$ by \eqref{eq:mc}, the increment
$\log(1 - \lambda + \lambda E_i)$ lies in $[\log(1-\lambda), \log(1+\lambda M)]$
for $\lambda < 1$, whatever the score; for $\lambda = 1$ and $|s_i| \le B$,
every term of \eqref{eq:mc} lies between $e^{-B}$ and $e^{B}$, so
$\log E_i \in [-2B, \log(M+1)]$. In both cases the differences
$\log(1 - \lambda + \lambda E_i) - g_{M,i}(\lambda)$ are bounded martingale
differences for the filtration $(\G_{i+1})$, and the martingale strong law
(Azuma--Hoeffding and Borel--Cantelli, as in the proof of
Proposition~\ref{prop:necessary}) shows that their average tends to zero
almost surely.
\end{proof}

Part (i) needs no assumption on the scores: for $\lambda < 1$ each increment
$\log(1-\lambda+\lambda E_i)$ lies in $[\log(1-\lambda), \log(1+\lambda M)]$,
whatever the learner. Since $1-\lambda+\lambda E$ is a weighted average of 1
and $E$, concavity of the logarithm gives $\log(1-\lambda+\lambda E) \ge \lambda\log E$,
so $g_{M,i}(\lambda) \ge \lambda\, g_{M,i}$: betting a fraction $\lambda$ keeps
at least a fraction $\lambda$ of the growth rate. Part (ii) needs
$|s_t| \le B$, which gives $\log E_i \ge -2B$; clipping the score keeps
validity (Theorem~\ref{thm:valid}), and our learners clip at $B = 30$. A ridge
penalty alone does not bound the score: the estimate can grow like
$\sqrt t$, and polynomial features are unbounded.

The same theory gives the rate of false alarms of a twin that ignores the
action. Theorem~\ref{thm:growth} depends only on the laws of the real outcome
and of the simulations, which for such a twin are $P_a$ and $\bar P$ even
under the true null; the monitor therefore grows as if it were testing $P_a$
against $\bar P$.

\begin{corollary}[False-alarm rate of an action-ignoring twin]\label{cor:falsealarm}
In the setting of Proposition~\ref{prop:necessary}, suppose that the actions
take finitely many values, that the laws $P_a$ and $\bar P$ and the score
$T_a = e^{s_a}$ used when $a$ is played, with $|s_a| \le B$, do not change
over time, that $\KL(P_a\|\bar P) < \infty$ for every action $a$, and that the
fraction of the first $t$ steps at which the policy plays $a$ converges almost
surely to $\varpi_a$. Let $g_M(a) = g_M(P_a, \bar P, T_a)$ be the growth rate
\eqref{eq:growthrate} with the real law $P_a$ in the role of $Q$ and the
simulator law $\bar P$ in the role of $P$, $g_\infty(a)$ its limit as
$M \to \infty$, and
$\tilde T_a = T_a\bar P/\E_{\bar P}[T_a]$. Then, under $H_0$:
\begin{enumerate}
\item[(i)] $g_\infty(a) = \KL(P_a\|\bar P) - \KL(P_a\|\tilde T_a)$ and
$g_M(a) \ge g_\infty(a) - (\rho_a - 1)/(M+1)$, where
$\rho_a = \E_{P_a}[T_a]/\E_{\bar P}[T_a]$ is the limit of
Proposition~\ref{prop:necessary}(i). If $1 \in \Lambda$, almost surely
$\liminf_t t^{-1}\log W_t \ge \sum_a \varpi_a\, g_M(a)$.
\item[(ii)] For every $\lambda$, every $M$ and every score,
$g_M(\lambda; P_a, \bar P, T_a) \le \KL(P_a\|\bar P)
\le \log\{1 + \chi^2(P_a\|\bar P)\}$, and almost surely
$\limsup_t t^{-1}\log W_t \le \sum_a \varpi_a \KL(P_a\|\bar P)$. The
likelihood-ratio score attains the bound as $M \to \infty$.
\item[(iii)] If the twin is run under randomised actions,
$\bar P = \sum_b u_b P_b$ with $u_b > 0$, then
$\KL(P_a\|\bar P) \le \log(1/u_a)$ and $\chi^2(P_a\|\bar P) \le 1/u_a - 1$;
with $K$ actions drawn uniformly, $\log K$ and $K - 1$.
\end{enumerate}
\end{corollary}

The proof is in Appendix~\ref{app:proofs}. The false alarm therefore arrives
after roughly $\log(1/\alpha)/\sum_a \varpi_a g_M(a)$ steps: the rate is a
Kullback--Leibler divergence, not the $\chi^2$ divergence that governs the
expected bet, and it is largest for a policy that concentrates on one action,
$\varpi_a \to 1$, as Thompson sampling does. The rate vanishes only if the arms
have the same law, and with $K$ arms it can never exceed $\log K$ per step.

\section{Betting is predicting}\label{sec:predictive}

By Theorem~\ref{thm:growth}, a good score is one whose tilted null is close
to the truth. Choosing such a score is therefore a problem of
sequential prediction: before each real outcome is seen, forecast it as well
as possible from the past. When the null density is known, this problem has a
classical solution, the Bayes factor, and we build on it. We first recall why
the Bayes factor is a product of forecasts and show that the Monte Carlo
e-process has the same structure, with the score in place of the model
(Section~\ref{sec:scores}); we then derive an update rule for the score with a
pathwise guarantee (Section~\ref{sec:simlik}), and its practical version
(Section~\ref{sec:blr}).

For a model $q_\theta$ with prior $\pi_0$, the Bayes factor is the ratio of
the marginal likelihood $m(x_{1:t}) = \int \prod_{i \le t} q_\theta(x_i \mid x_{1:i-1})\,
\pi_0(d\theta)$ to the null density $p_0$. By the chain rule,
\begin{equation}\label{eq:prequential}
\mathrm{BF}_t = \prod_{i=1}^{t} \frac{m_i(x_i)}{p_0(x_i \mid x_{1:i-1})},
\qquad
m_i(x) = \int q_\theta(x \mid x_{1:i-1})\,\pi_{i-1}(d\theta),
\end{equation}
where $\pi_{i-1}$ is the posterior after $i-1$ observations, so $m_i$ is the
posterior predictive of $x_i$. (For a coin with a uniform prior,
$m_i(\text{heads}) = (k+1)/(i+1)$ after $k$ heads in $i-1$ tosses.) Each
factor has conditional mean $\int m_i = 1$ under the null, whatever the prior;
$\log \mathrm{BF}_t$ is the difference between the cumulative log scores of
the learner and of the null \citep{dawid1984present}; and for i.i.d.\ data
and a regular model with $d$ parameters
\[
\log \mathrm{BF}_t = \max_\theta \sum_{i \le t} \log\frac{q_\theta(x_i)}{p_0(x_i)}
- \frac d2 \log t + O_P(1)
\]
\citep{clarke1990information}. The Monte Carlo e-process has the same
structure, and we now look at it from this point of view: the score, through
its tilted null, plays the role of the forecast $m_i$, the simulations that of
the null density $p_0$, and a family of scores $T_\theta$ that of the model
$q_\theta$.

\subsection{Scores are forecasts}\label{sec:scores}

Fix a time $t$ and condition on $\G_t$, so that the action is known. Write
$p_t$ for the density of the simulator's next outcome under that action, and
$q_t$ for the density of the unknown law $Q_t$ of the real next outcome. The
natural score is the density ratio $q_t/p_t$. Neither density can be
evaluated, but their ratio can be estimated by a classifier separating real
outcomes (label 1) from simulator draws (label 0): if the two classes have
prior weights $r$ and $1-r$, Bayes' theorem gives
\[
\Pr(\text{real} \mid x) = \frac{r\, q_t(x)}{r\, q_t(x) + (1-r)\, p_t(x)},
\qquad
\log\frac{\Pr(\text{real} \mid x)}{\Pr(\text{simulated} \mid x)}
= \log\frac{q_t(x)}{p_t(x)} + \log\frac{r}{1-r},
\]
so the classifier's logit is the log ratio up to an additive constant, which
is zero with balanced class weights \citep{thomas2022lfire} and in any case
cancels in \eqref{eq:mc}.

In practice $T_t$ is only an estimate of the true ratio, learned from past
data, and what matters is the density it induces. Reweighting the simulator by
the score as in \eqref{eq:tilt}, which keeps the simulator's outcomes and gives
each a weight proportional to its score, yields the tilted null
\[
\tilde T_t(x) = \frac{T_t(x)\,p_t(x)}{Z_t},
\qquad
Z_t = \E_{P_t}[T_t] = \int T_t(x)\,p_t(x)\,dx ,
\]
a density on the space of outcomes, not the distribution of the score values.
It is the alternative implied by the bet in the sense of
\citet{shafer2021testing}. As $M \to \infty$, $\widehat Z_t \to Z_t$ by the
law of large numbers, so the bet tends to
\[
E_t = \frac{T_t(X_t)}{\widehat Z_t} \;\to\; \frac{T_t(X_t)}{Z_t}
= \frac{\tilde T_t(X_t)}{p_t(X_t)} ,
\]
a likelihood ratio of $\tilde T_t$ against the simulator: a bettor using the
score acts as if the next outcome followed $\tilde T_t$, and wins when real
outcomes fall where $\tilde T_t$ puts more mass than the simulator. If $T_t$
is the true ratio, $Z_t = 1$ and $\tilde T_t = q_t$; otherwise $\tilde T_t$
is the score's forecast of $q_t$. Conversely, any predictive density $f_t$
computed from the past is the tilted null of a score,
\[
T_t = \frac{f_t}{p_t}
\quad\Longrightarrow\quad
Z_t = 1,
\qquad
\tilde T_t = f_t ,
\]
and multiplying a score by a constant $c > 0$ changes neither the bet
\eqref{eq:mc}, where $c$ cancels, nor its tilted null. Scores up to a
constant and predictive densities are therefore the same object. The
correspondence is the familiar one between density ratios and densities,
used for instance in noise-contrastive estimation \citep{gutmann2012noise};
what the Monte Carlo normalisation adds is that the wealth scores these
densities prequentially, without evaluating any of them.

Indeed, by the limit of the bet above, with the betting fraction
$\lambda = 1$ the log wealth tends to
\begin{equation}\label{eq:prequential-wealth}
\log W_t \;=\; \sum_{i \le t} \log E_i \;\to\; \sum_{i \le t} \log \tilde T_i(X_i)
- \sum_{i \le t} \log p_i(X_i),
\end{equation}
which has the form of $\log \mathrm{BF}_t$ in \eqref{eq:prequential}, with
the tilted null $\tilde T_i$ in place of the posterior predictive $m_i$ and
the simulator $p_i$ in place of the null density $p_0$. It is the difference
between the cumulative log scores of two forecasters, each forecast issued
before the outcome it is scored on: the prequential likelihood ratio of
\citet{dawid1984present}. The monitor rejects at level $\alpha$ when the
score's forecasts have predicted the real outcomes $1/\alpha$ times better
than the simulator did. Under the truth, writing
$\log(\tilde T_t/p_t) = \log(q_t/p_t) - \log(q_t/\tilde T_t)$, the expected
increment of \eqref{eq:prequential-wealth} is
\[
\E_{Q_t}\Big[\log\frac{\tilde T_t(X_t)}{p_t(X_t)}\Big]
= \KL(Q_t\|P_t) - \KL(Q_t\|\tilde T_t).
\]
The first term is the information against the simulator and does not depend
on the score; the second is the error of the score's forecast, which by
Gibbs' inequality is nonnegative and vanishes only when $\tilde T_t = Q_t$,
that is, when $T_t$ is the true ratio up to a constant. This is
Theorem~\ref{thm:growth}(ii), and a finite $M$ multiplies the growth by about
$M/(M+1)$ (Theorem~\ref{thm:growth}(iv)). A learned score therefore pays, at
every step, for its error as a forecaster, and only in power.

Reading the score as a forecast scored by the logarithmic loss turns the
choice of the score into a problem of sequential prediction, which
Section~\ref{sec:rules} solves with a Bayesian mixture. The reading also says
where power comes from. A score can only forecast within the family of its
tilted nulls, so the features determine which departures can be detected: for
a score built from the simulator's own parameters, as in SLICE, the tilted
nulls belong to that parametric family, and a departure outside it is
invisible however accurate the score (Section~\ref{sec:exp-concurrent}),
whereas a score learned on features of the outcome reweights the simulator in
directions that need not correspond to any parameter. And since a forecast
need only be issued before the outcome it is scored on, any learner, exact or
approximate, trained online or in advance, keeps the validity of
Theorem~\ref{thm:valid}.

\subsection{From Bayes to the simulation likelihood}\label{sec:rules}\label{sec:exactbayes}\label{sec:simlik}

An update rule for the score must fix the bet on
$X_t$ before $X_t$ is seen, should hedge early on, since an overconfident
forecast can lose all the growth (Theorem~\ref{thm:growth}), and must use
simulations alone, since $p_t$ can only be sampled.

Bayes' rule hedges by averaging over what has not yet been learned. If the
real outcomes followed a model $\{Q_\theta\}$ with prior $\pi_0$, the best
forecast under the model would be its posterior predictive $m_t$, since the
logarithmic score is strictly proper. Section~\ref{sec:scores} supplies a
model: the log-linear scores $s = \theta^\top\phi$ have tilted nulls
$dQ_\theta/dP = \exp\{\theta^\top\phi - \log Z(\theta)\}$, with
$Z(\theta) = \E_P[e^{\theta^\top\phi}]$, an exponential family built on the
simulator. With this model the forecast is an average of tilted nulls,
$m_t = \int q_\theta\,\pi_{t-1}(d\theta)$, its score is
$T_t = \int (dQ_\theta/dP)\,\pi_{t-1}(d\theta)$, and the process becomes a
Bayes factor, the benchmark for what follows.

\begin{proposition}[Predictive scores give the Bayes factor; cf.\ \citealp{clarke1990information}]\label{prop:bayes}
With this score $\E_P[T_t] = 1$, the unlimited-budget factor is
$E_t^\infty = T_t(X_t)$, and
\[
\prod_{i\le t} E_i^\infty = \int \prod_{i \le t} \frac{dQ_\theta}{dP}(X_i)\,\pi_0(d\theta),
\]
the Bayes factor of the tilted family against the simulator. For i.i.d.\ data
and under the usual regularity conditions for Laplace approximations, its
logarithm is $\sup_\theta \sum_{i\le t}\log (dQ_\theta/dP)(X_i) - \frac{d}{2}\log t
+ O_P(1)$, and $t^{-1}$ times it converges to
$\KL(Q\|P) - \inf_\theta \KL(Q\|Q_\theta)$.
\end{proposition}

The proof, in Appendix~\ref{app:proofs}, combines the chain rule with the
Laplace expansion of \citet{clarke1990information}.

Proposition~\ref{prop:bayes} is only a benchmark: Bayes' rule needs the
normalising constant $Z(\theta)$ of every tilt at every step, an integral
against a simulator that can only be sampled.

The Monte Carlo factor provides the way out, because it estimates that
constant from the $M+1$ outcomes. Fix $\theta \in \mathbb{R}^d$ and
$\G_t$-measurable features $\phi_t$, write $U_{t,0} = X_t$ and
$U_{t,k} = Y_{t,k}$, and let
\[
E_t(\theta) = \frac{(M+1)\exp\{\theta^\top\phi_t(U_{t,0})\}}
{\sum_{k=0}^{M}\exp\{\theta^\top\phi_t(U_{t,k})\}}
\]
be the factor \eqref{eq:mc} with score $s_t = \theta^\top\phi_t$. As a
function of $\theta$ it is, up to the constant $M+1$, the probability that the
score assigns to the real outcome among the $M+1$
(Section~\ref{sec:method}), a conditional-logit likelihood whose normalising
constant is a sum over the simulations, always computable. We call
$L_t(\theta) = \prod_{i \le t} E_i(\theta)$ the \emph{simulation likelihood};
its logarithm is the InfoNCE objective of contrastive learning
\citep{oord2018representation,ma2018noise}, which \citet{durkan2020contrastive}
use to estimate likelihood ratios in simulation-based inference. With a prior $\pi_0$ on $\theta$
and posterior $\pi_t \propto \pi_0 L_t$, the \emph{posterior-predictive bet}
is
\[
\bar E_t = \int E_t(\theta)\, \pi_{t-1}(d\theta),
\]
updated by one posterior step per observation instead of a refit on the whole
history. The posterior is a prequential posterior in the sense of
\citet{sinharoy2026prequential}, with loss $-\log E_t(\theta)$ and learning
rate one, the choice that makes the posterior mixture of the bets an
e-process.

\begin{theorem}[Predictive updating on the simulation likelihood]\label{thm:simlik}
Suppose the conditions of Theorem~\ref{thm:valid} hold for each fixed $\theta$
(simulations under the action taken, $\phi_t$ $\G_t$-measurable), that
$\|\phi_t\| \le G$, and that $\pi_0$ is the uniform distribution on a convex
compact set $\Theta \subset \mathbb{R}^d$ with nonempty interior and diameter
at most $2D$.
\begin{enumerate}
  \item[(i)] (Exact factorisation and validity.) The wealth
        $\bar W_t = \prod_{i \le t} \bar E_i$ equals
        $\int L_t(\theta)\,\pi_0(d\theta)$. It is a test martingale under
        $H_0$, for every policy, and it is updated by one posterior step per
        observation, with no recomputation.
  \item[(ii)] (Pathwise regret.) On every sample path, for every
        $t \ge d/(4GD)$,
        \[
        \log \bar W_t \;\ge\; \sup_{\theta \in \Theta} \log L_t(\theta)
        - d \log t - d \log\frac{4eGD}{d}.
        \]
        If the maximiser $\theta^*_t$ of $L_t$ lies in $\Theta$ at distance at
        least $r_t = \sqrt{d/(tG^2)}$ from its boundary, the bound improves to
        $\log \bar W_t \ge \log L_t(\theta_t^*) - \frac d2 \log t
        - \frac d2 \log\frac{e G^2 (2D)^2}{d}$.
  \item[(iii)] (Growth.) Let $g_{M,i}(\theta) = \E[\log E_i(\theta) \mid \G_i]$,
        the conditional growth rate \eqref{eq:growthrate} of the fixed score
        $T = e^{\theta^\top\phi}$, and $g_M(\theta) = g_M(Q, P, e^{\theta^\top\phi})$.
        For every $\theta \in \Theta$, almost surely,
        \[
        \liminf_{t \to \infty} \frac1t \Big(\log \bar W_t - \sum_{i \le t} g_{M,i}(\theta)\Big) \ge 0 .
        \]
        In particular, for i.i.d.\ data under a fixed action and features,
        $\liminf_t t^{-1}\log \bar W_t \ge \sup_{\theta\in\Theta} g_M(\theta)$,
        and, if $\KL(Q\|P) < \infty$, by Theorem~\ref{thm:growth}
        $g_M(\theta) \ge \KL(Q\|P) - \KL(Q\|Q_\theta) - \log\{1 + (\rho_\theta - 1)/(M+1)\}$,
        where $Q_\theta$ is the tilt of $P$ by $\theta^\top\phi$ and
        $\rho_\theta = \E_Q[e^{\theta^\top\phi}]/\E_P[e^{\theta^\top\phi}]$.
\end{enumerate}
\end{theorem}

Part (i) says that the score can be updated by Bayes' rule exactly, even
though the normalising constant $Z(\theta)$ of the tilt is unknown. This is
because the simulation likelihood is not the density of the real outcome under
the tilt, which would need $Z(\theta)$, an integral over all possible
outcomes. It answers the simpler question of Section~\ref{sec:method}: among
the $M+1$ outcomes at hand, which one is the real one? The answer compares the
scores of these $M+1$ outcomes only, so it needs no integral, and its
posterior predictive is the bet itself: updating $\theta$ by Bayes' rule and
betting are the same operation, so the guarantees of Bayesian mixtures hold
for the wealth. It uses less information, and as $M \to \infty$ it recovers
the Bayes factor of Proposition~\ref{prop:bayes}.

Part (ii) holds on every path, with no assumption on the alternative: the
monitor is never more than $d\log t$ behind the best fixed score in the class
chosen in hindsight and evaluated on the same simulations. The argument is the
standard one for Bayesian mixtures of log-concave likelihoods
\citep{kakade2005online,grunwald2007minimum}, applied to the simulation
likelihood, whose log is concave in $\theta$ and whose gradient is bounded by
$2G$. Part (iii) turns this into growth at the rate of the best tilt in the
class, up to the simulation cost of Theorem~\ref{thm:growth}(iii), and
Ville's inequality then gives rejection after about
$\{\log(1/\alpha) + d\log t\}/\sup_\theta g_M(\theta)$ steps.

\subsection{Practical update rules}\label{sec:blr}\label{sec:algorithm}

Exact Bayesian updating needs the posterior $\pi_{t-1}$ at every step. We
replace it by a Gaussian approximation updated in one step per observation,
on either of two losses, or update the forecast directly as martingale
posteriors do. All give valid bets; they differ in where uncertainty comes from
and in the guarantees they carry (Table~\ref{tab:rules}).

The integral in $\bar E_t$ is over $\theta$ only, and any $\G_t$-measurable
distribution $\nu_{t-1}$ on $\theta$, an approximate posterior, in place of
$\pi_{t-1}$ still gives a valid bet
$E^{\nu}_t = \int E_t(\theta)\,\nu_{t-1}(d\theta)$, because $E_t(\theta)$ is
fair for each $\theta$. The price of the approximation is bounded as follows.

\begin{proposition}[Approximate predictive bets]\label{prop:approx}
Let $\nu_{t-1}$ be any $\G_t$-measurable distribution on $\mathbb{R}^d$ with mean
$\bar\theta_{t-1}$ and covariance $\Sigma_{t-1}$, and $\|\phi_t\| \le G$. Then
$E^{\nu}_t$ is an e-value given $\G_t$, and
\[
\log E^{\nu}_t \;\ge\; \E_{\nu_{t-1}}\big[\log E_t(\theta)\big]
\;\ge\; \log E_t(\bar\theta_{t-1}) - \frac{G^2}{2}\,\mathrm{tr}\,\Sigma_{t-1}.
\]
\end{proposition}

\begin{proof}
Validity follows from Fubini's theorem. The first inequality is Jensen's.
The function $\theta \mapsto \log E_t(\theta)$ is concave, with gradient
$\phi_t(U_{t,0}) - \sum_k w_k \phi_t(U_{t,k})$ and Hessian
$-\mathrm{Cov}_w(\phi_t)$, where $w_k \propto e^{\theta^\top\phi_t(U_{t,k})}$;
since $\mathrm{Cov}_w(\phi_t) \preceq \E_w[\phi_t\phi_t^\top] \preceq G^2 I$, with $\E_w$
the average over $k$ of $\phi_t(U_{t,k})$ with weights $w_k$,
Taylor's theorem with integral remainder around $m = \bar\theta_{t-1}$ gives
$\log E_t(\theta) \ge \log E_t(m) + \nabla\log E_t(m)^\top(\theta - m)
- \frac{G^2}{2}\|\theta - m\|^2$ for every $\theta$. Taking the expectation
under $\nu_{t-1}$, the linear term vanishes because $m$ is its mean, and
$\E_{\nu_{t-1}}\|\theta - m\|^2 = \mathrm{tr}\,\Sigma_{t-1}$.
\end{proof}

A predictive bet built from an approximate posterior is therefore never worse
than the plug-in bet at its mean by more than
$\frac{G^2}{2}\sum_i \mathrm{tr}\,\Sigma_{i-1}$. If the posterior precision grows
at least linearly, $\Sigma_i^{-1} \succeq \Sigma_0^{-1} + i\,c_0 I$ for some
$c_0 > 0$, this is at most $\frac{G^2 d}{2c_0}\log t + O(1)$, the order of the
regret in Theorem~\ref{thm:simlik}; such growth needs features that are not
collinear under the simulator and iterates that stay bounded. The Bayesian
learning rule \citep{khan2023bayesian} provides such a posterior. It
approximates a posterior within an exponential family by natural-gradient
descent on the expected loss plus the Kullback--Leibler divergence from the
prior; in a stream, one step per observation with the previous approximation
as the prior \citep{jones2024bong} gives, for a Gaussian
$\nu_t = \mathcal N(\bar\theta_t, S_t^{-1})$ and a per-step loss $\ell_t$,
\begin{equation}\label{eq:blr}
S_t = S_{t-1} + \E_{\nu_{t-1}}[\nabla^2 \ell_t], \qquad
\bar\theta_t = \bar\theta_{t-1} - S_t^{-1}\, \E_{\nu_{t-1}}[\nabla \ell_t],
\end{equation}
which is exact Bayesian updating (the Kalman filter) for a Gaussian likelihood
and one step of online Newton in general. Its cost is $O((M+1)d^2 + d^3)$ per
step with a direct solve, $O((M+1)d^2)$ with rank-one updates of $S_t^{-1}$,
with no refitting and no burn-in.

Two losses are natural. On the simulation likelihood,
$\ell_t(\theta) = -\log E_t(\theta)$, with gradient
$\sum_k w_k \phi_t(U_{t,k}) - \phi_t(U_{t,0})$ and Hessian
$\mathrm{Cov}_w(\phi_t)$; expectations under $\nu_{t-1}$ are computed with
fixed antithetic nodes, and the bet is the average of $E_t(\theta)$ over the
nodes, so Theorem~\ref{thm:simlik} and Proposition~\ref{prop:approx} apply
directly. Alternatively, $\ell_t$ is the balanced logistic loss of classifying
the real outcome (weight one) against the $M$ simulations (weight $1/M$ each),
a generalised posterior \citep{bissiri2016general,sinharoy2026prequential}
that identifies the same tilt through noise-contrastive estimation
\citep{gutmann2012noise}. With the probit approximation
$\E_\nu[\sigma(z^\top\theta)] \approx \sigma(\kappa\bar\eta)$, where
$\bar\eta = z^\top\bar\theta$, $v = z^\top S^{-1}z$ and
$\kappa = (1+\pi v/8)^{-1/2}$, the score is the predictive log-odds
$\kappa\bar\eta$ on the probability scale, inserted in \eqref{eq:mc}; this is
our default. The ratio scale $\bar\eta + v/2$ would be the exact predictive
under an exact posterior, but its premium $v/2$ over-bets where the
approximation is least reliable, and it is the worst score in every experiment
(Supplement, Section~S4). Both losses give scores that start at zero, so the
monitor starts by not betting, and that shrink where the posterior is
uncertain. Features are standardised with simulator draws only, before any
real data are seen, so that the score stays predictable.

The same two classifiers can also be updated without a posterior over their
weights, by updating their predictive directly. In the approach of
\citet{fong2023martingale} one specifies a rule
$m_{t+1} = \mathcal U_t(m_t, X_t)$ and recovers uncertainty by
\emph{predictive resampling}, drawing future outcomes from the current
predictive and updating; this converges when imputing one more outcome does
not move the forecast on average, $\E_{X \sim m_t}[\mathcal U_t(m_t, X)] = m_t$,
the conditionally identically distributed sequences of \citet{berti2004limit}.
Here the learner carries the predictive itself, $\sigma(z^\top\theta_t)$ for
the logistic loss or the softmax $e^{\theta_t^\top z_k}/\sum_{k'} e^{\theta_t^\top z_{k'}}$
over the $M+1$ outcomes for the simulation likelihood, and updates it by
Fisher scoring on the predictive's own log-likelihood,
\begin{equation}\label{eq:mp}
I_t = I_{t-1} + J_t(\theta_{t-1}), \qquad
\theta_t = \theta_{t-1} - I_t^{-1}\, \nabla_t(\theta_{t-1}),
\end{equation}
where $\nabla_t$ is the gradient of the new batch's loss, $J_t$ its Fisher
information, which for both losses does not depend on the labels, and $I_0$ a
prior precision. Under the predictive the gradient has mean zero, so
$\theta_t$ is a martingale, and the bet averages the factor over $N$
predictive-resampling paths, obtained by running \eqref{eq:mp} forward from
$(\theta_{t-1}, I_{t-1})$ on past batches whose labels are drawn from the
predictive ($N = 16$, truncated at $\min(t, 200)$ steps, refreshed every ten
steps). Any learner can be made a martingale-posterior learner in this way:
with the empirical predictive, predictive resampling is the Bayesian
bootstrap, and a random forest fitted with Dirichlet weights on the past steps
is valid as well (Supplement, Section~S4).

\begin{table}[t]
\centering
\caption{Update rules for the score (Sections~\ref{sec:simlik}
and~\ref{sec:blr}). All give exactly valid bets; they differ in the source of
uncertainty about the score and in the guarantee they carry beyond validity.
Further variants are in the Supplement.}
\label{tab:rules}
\small
\begin{tabular}{lll}
\toprule
rule & uncertainty from & guarantee beyond validity \\
\midrule
Bayes on the simulation likelihood & exact posterior & pathwise regret (Theorem~\ref{thm:simlik}) \\
BLR on the simulation likelihood & Gaussian approximation & Theorem~\ref{thm:simlik} up to Proposition~\ref{prop:approx} \\
BLR on the logistic loss (default) & Gaussian approximation & none \\
predictive-first \eqref{eq:mp} & predictive resampling & none \\
\bottomrule
\end{tabular}
\end{table}

\begin{figure}[t]
\centering
\includegraphics[width=\textwidth]{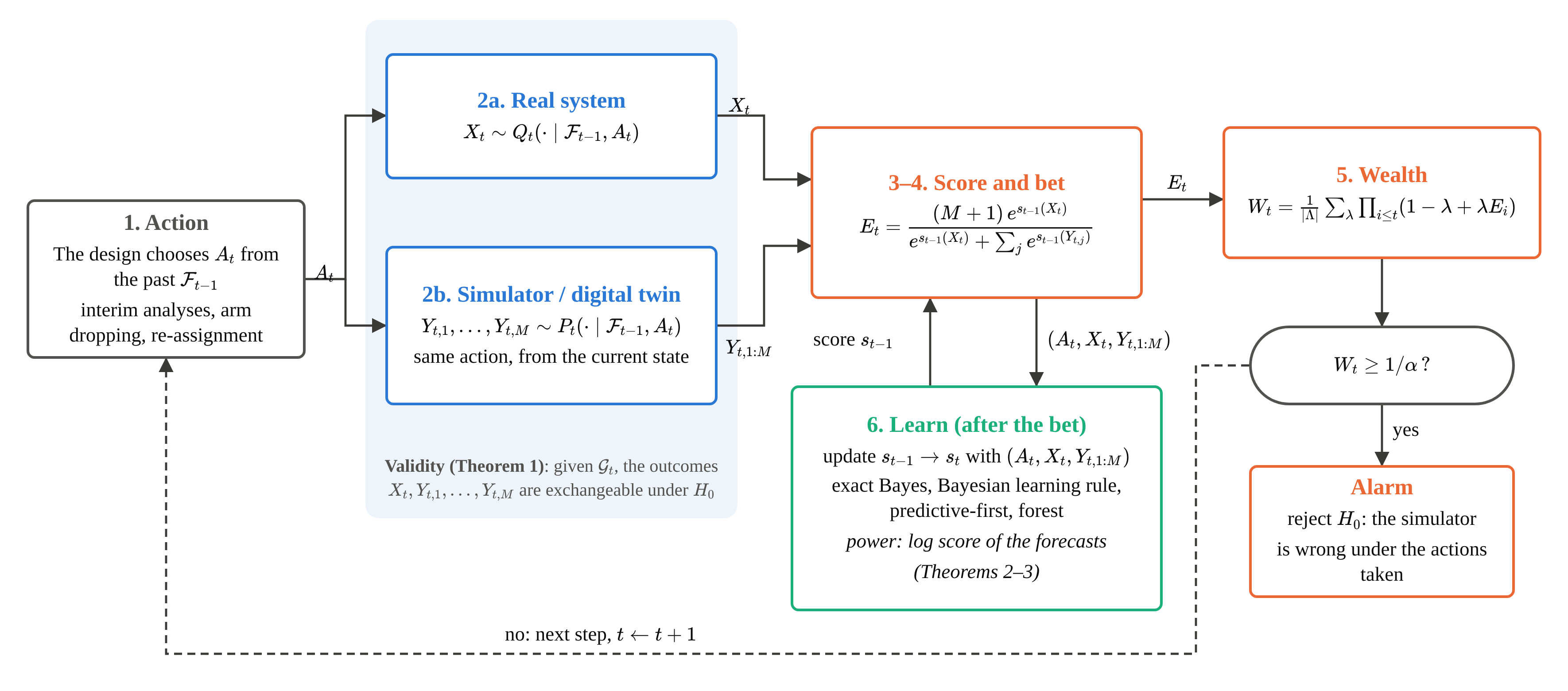}
\caption{One step of the procedure. The design chooses the action from the
past (1); the real system and the simulator are run under that action and
from the current state (2a, 2b); the current score, fixed before $X_t$ is
seen, turns the $M+1$ outcomes into the bet $E_t$ (3--4), which updates the
wealth (5); the null is rejected when the wealth reaches $1/\alpha$, and
otherwise the learner updates the score with the new outcomes (6) and the
next step begins. Validity rests on the shaded block alone; the learner
affects only power. Algorithm~1 gives the
formulas of each step.}
\label{fig:workflow}
\end{figure}

Algorithm~1 puts the default rule together, the logistic Bayesian learning
rule with the probability-scale predictive; its steps are numbered as in
Figure~\ref{fig:workflow}, and each is followed by the role it plays.

\paragraph{Algorithm 1 (Monte Carlo e-process with a predictive score).}
\textbf{Input:} a simulator that can be run under any action from any state; a
feature map $z(x)$; the number of simulations $M$; the level $\alpha$; the grid
$\Lambda$ of betting fractions. \textbf{Initialise:} standardise $z$ with
simulator draws only; set $\bar\theta_0 = 0$, $S_0 = I$ (vaguer on the intercept), and
$W_0 = 1$. \textbf{Notation:} $U_{t,0} = X_t$, $U_{t,j} = Y_{t,j}$,
$z_k = z(U_{t,k})$; $s_{t-1}$ is the score fixed at the end of step $t-1$, the
$s_t$ of \eqref{eq:mc}.
\begin{enumerate}
  \item \textbf{Action.} Observe the action $A_t$ chosen by the design from the
        past (arm, prompt, dose). --- It defines the null law $P_t$.
  \item \textbf{Simulate.} Draw $Y_{t,1}, \ldots, Y_{t,M}$ from the simulator
        under $A_t$ from the current state, and observe $X_t$.
        --- Exchangeable with $X_t$ under $H_0$.
  \item \textbf{Score.} For each of the $M+1$ outcomes $x$, compute
        \[
        \begin{gathered}
        \bar\eta = z(x)^\top \bar\theta_{t-1}, \qquad v = z(x)^\top S_{t-1}^{-1} z(x), \\
        \kappa = (1 + \pi v/8)^{-1/2}, \qquad
        s_{t-1}(x) = \mathrm{clip}(\kappa\,\bar\eta,\, -B,\, B) .
        \end{gathered}
        \]
        --- $s_{t-1}$ is the predictive log-odds of ``real'' against
        ``simulated'', an estimate of $\log(q/p)$ shrunk where the posterior is
        uncertain; it forecasts $\tilde T_t \propto p_t\, e^{s_{t-1}}$
        \eqref{eq:tilt}. Clipping at $B$ (we use $B = 30$) keeps the score
        predictable, so validity is unaffected, and makes
        Corollary~\ref{cor:cumulative}(ii) apply.
  \item \textbf{Bet.}
        \[
        E_t = \frac{(M+1)\, e^{s_{t-1}(X_t)}}{e^{s_{t-1}(X_t)} + \sum_{j=1}^{M} e^{s_{t-1}(Y_{t,j})}} .
        \]
        --- Exactly fair under $H_0$ (Theorem~\ref{thm:valid}); close to
        $\tilde T_t(X_t)/p_t(X_t)$ for large $M$.
  \item \textbf{Wealth.} $W_t = |\Lambda|^{-1}\sum_{\lambda\in\Lambda}
        \prod_{i\le t}(1-\lambda+\lambda E_i)$; reject $H_0$ when
        $W_t \ge 1/\alpha$. --- A test martingale; its log is the
        prequential likelihood ratio \eqref{eq:prequential-wealth} of the
        forecasts against the simulator, and it grows at about
        $\frac{M}{M+1}\{\KL(Q\|P) - \KL(Q\|\tilde T)\}$ per step
        (Theorem~\ref{thm:growth}).
  \item \textbf{Learn.} Only now, with labels $y_0 = 1$ for $X_t$ and $y_j = 0$
        for $Y_{t,j}$, weights $w_0 = 1$ and $w_j = 1/M$, and
        $p_k = \sigma(s_{t-1}(U_{t,k}))$, update
        \[
        \begin{gathered}
        \nabla_t = \sum_{k=0}^{M} w_k (p_k - y_k)\, z_k, \qquad
        J_t = \sum_{k=0}^{M} w_k \kappa_k\, p_k (1 - p_k)\, z_k z_k^\top, \\
        S_t = S_{t-1} + J_t, \qquad \bar\theta_t = \bar\theta_{t-1} - S_t^{-1} \nabla_t .
        \end{gathered}
        \]
        --- One step of the Bayesian learning rule \eqref{eq:blr}, with $\kappa_k$
        as in step 3 for $z_k$: $\kappa_k p_k(1-p_k)$ is the derivative of the
        probit approximation $\sigma(\kappa_k\bar\eta_k)$ with respect to
        $\bar\eta_k$, which approximates $\E_\nu[\sigma'(z_k^\top\theta)]$; the score
        stays predictable.
\end{enumerate}
\textbf{Variants.} With the simulation likelihood, steps 3--4 are replaced by
the posterior-predictive mixture $\bar E_t = N^{-1}\sum_b E_t(\theta_b)$ over
posterior nodes $\theta_b$, and step 6 by \eqref{eq:blr} with
$\ell_t = -\log E_t(\theta)$; Theorem~\ref{thm:simlik} and
Proposition~\ref{prop:approx} then apply. With the predictive-first rule,
steps 3 and 6 are replaced by \eqref{eq:mp} and step 4 by the average of the
factor over predictive-resampling paths. In a multi-arm design, one copy of the
algorithm per arm, betting only on that arm's observations, is combined by e-BH
at the end (Section~\ref{sec:trial}).

\section{Simulation studies}\label{sec:experiments}

The experiments ask whether a learned score can break validity
(Section~\ref{sec:exp-drift}); whether adaptive allocation can
(Section~\ref{sec:exp-bandit}); whether the growth rates of
Section~\ref{sec:power} and the regret of Theorem~\ref{thm:simlik} are what
one observes (Section~\ref{sec:exp-budget}); how the update rules and the
scores of concurrent methods compare (Sections~\ref{sec:exp-blr}
and~\ref{sec:exp-concurrent}); and what adaptive actions add in an audit with
real model outputs (Section~\ref{sec:exp-llm}). Throughout, $\alpha = 0.05$,
$M = 16$ unless stated otherwise, and the betting fraction is mixed over 20
values from 0.05 to 1. Each table reports, over 400 replications on data
shared by all methods, the share of runs that reject a true null within the
horizon (false alarms; Monte Carlo standard error about 0.011 near 0.05), the
share that reject a false null (power), and the median stopping time, with
runs that never reject counted as infinite; the horizon is 500 steps for the
drift task and 400 for the bandit and the audit.

\subsection{Validity with learned scores}\label{sec:exp-drift}

The simulator is the two-dimensional ``banana''
$x = (z_1, z_2 + b(z_1^2 - 1))$, $z \sim \mathcal N(0, I)$, with $b = 0.5$; under
the alternative the real data drift to $b = 0.8$, which changes neither the
means nor the marginal of $x_1$, so the shape of the discrepancy must be
learned. The score is a balanced polynomial logistic regression refitted every
ten steps, regularised (degree 4, ridge 1) or overfitting (degree 6, ridge
$10^{-3}$). With the same score, using the learned ratio directly as the bet
gives 10\% false alarms with the regularised learner and 59\% with the
overfitting one, while the Monte Carlo e-process gives 3\% with both
(Table~\ref{tab:drift}): a worse learner costs power, not validity. A
Kolmogorov--Smirnov test recomputed every ten steps gives 32\%. The oracle
likelihood-ratio e-process, unavailable in practice, stops after 35 steps;
the gap to the 198 steps of the learned score is the price of learning the
discrepancy online.

\begin{table}[t]
\centering
\caption{Generative-model drift (Section~\ref{sec:exp-drift}). The KS test is
applied per coordinate against a pool of simulator draws, with a Bonferroni
correction, and recomputed every ten steps.}
\label{tab:drift}
\begin{tabular}{llrrr}
\toprule
learner & method & false alarms & power & median stop \\
\midrule
regularised & MC e-process & 0.030 & 0.990 & 198 \\
 & plug-in & 0.102 & 0.995 & 158 \\
flexible & MC e-process & 0.028 & 0.838 & 320 \\
 & plug-in & 0.588 & 0.998 & 66 \\
-- & KS, recomputed & 0.323 & 0.750 & 250 \\
-- & oracle LR & 0.030 & 1.000 & 35 \\
\bottomrule
\end{tabular}
\end{table}

\subsection{Adaptive allocation}\label{sec:exp-bandit}

Three arms are allocated by Thompson sampling on the real outcomes, and a twin
simulates any arm, $y = \mu_k + e^{\epsilon/2} - e^{1/8}$ with
$\epsilon \sim \mathcal N(0, 1)$ and $\mu = (0, 0.2, 0.4)$. Under the
alternative arm~0 is better in reality than the twin predicts (mean shifted by
0.5), and Thompson sampling pulls it in 65\% of steps instead of 7\%. The
conditioned e-process stays valid, with 2.8\% false alarms
(Table~\ref{tab:bandit}), whereas the same e-process with the twin run on
uniformly random arms, the setting of Proposition~\ref{prop:necessary}, rejects
the true null in every run, and a per-arm $t$-test recomputed every ten steps
in 39\% of runs.

\begin{table}[t]
\centering
\caption{Thompson-sampling bandit against a digital twin
(Section~\ref{sec:exp-bandit}).}
\label{tab:bandit}
\begin{tabular}{lrrr}
\toprule
method & false alarms & power & median stop \\
\midrule
MC e-process, twin run on the chosen arm & 0.028 & 1.000 & 49 \\
MC e-process, twin run on random arms & 1.000 & 1.000 & 40 \\
plug-in & 0.055 & 1.000 & 44 \\
per-arm $t$-test, recomputed & 0.388 & 1.000 & 30 \\
oracle LR & 0.043 & 1.000 & 15 \\
\bottomrule
\end{tabular}
\end{table}

\subsection{Empirical growth rates}\label{sec:exp-budget}

\begin{figure}[t]
\centering
\includegraphics[width=\textwidth]{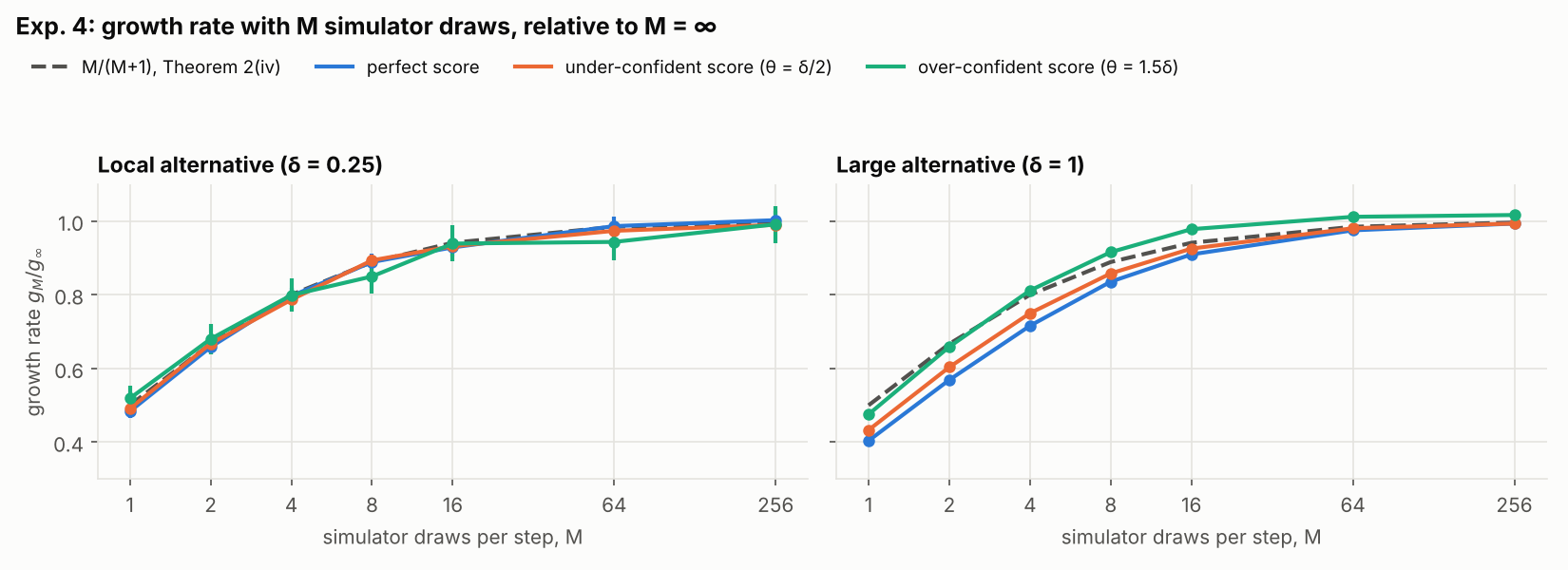}
\caption{Growth rate $g_M$ relative to $g_\infty$ for $P = \mathcal N(0,1)$,
$Q = \mathcal N(\delta, 1)$ and score $T(x) = e^{\theta x}$, estimated from
$4 \times 10^5$ draws (two standard errors shown). Dashed: $M/(M+1)$ from
Theorem~\ref{thm:growth}(iv).}
\label{fig:growth}
\end{figure}

On the drift task, validity holds at every budget, including $M = 1$ (false
alarms 1.5--3.5\%), and the median stopping time at $M = 1$ is 1.8 times that
at $M = 64$, close to the factor 2 of Theorem~\ref{thm:growth}(iv). With
$P = \mathcal N(0,1)$, $Q = \mathcal N(\delta, 1)$ and $T(x) = e^{\theta x}$,
where everything is explicit, the bounds of Theorem~\ref{thm:growth} hold in
all 56 configurations, and the efficiency $g_M/g_\infty$ follows $M/(M+1)$
closely for perfect, under- and overconfident scores alike
(Figure~\ref{fig:growth}). The regret bound of Theorem~\ref{thm:simlik},
checked by computing the mixture exactly on a grid for four scenarios
including a change point, holds on every path and at every time, with largest
regret 7.5 against a bound of 21.5 at $t = 2000$ and slopes close to the
interior rate $(d/2)\log t$ (Supplement, Section~S4).

\subsection{Update rules}\label{sec:exp-blr}

Table~\ref{tab:blr} compares the refitted point estimate with three update
rules of Section~\ref{sec:predictive}, updated at every step with no burn-in.
All are valid, whereas using the logistic posterior odds directly as the bet
gives 14--74\% false alarms. The logistic Bayesian learning rule is the
fastest and makes the method insensitive to the flexibility of the features
(power 0.97 with degree-6 features, against 0.84 for the refit). The
simulation-likelihood learner, the one with the guarantee of
Theorem~\ref{thm:simlik}, is about 6\% slower on the drift task. The bound of
Theorem~\ref{thm:simlik} holds on every path, and the Bayesian mixture pays for it with cautious early bets; the logistic learner bets more
aggressively, with no such bound, and on these tasks this pays. All variants
are in the Supplement (Section~S4), and a comparison with the Besag--Clifford
e-process of \citet{dombowsky2026besag} in Section~S3.

\begin{table}[t]
\centering
\caption{Update rules (Section~\ref{sec:blr}), MC e-process throughout. BLR
on the simulation likelihood uses the posterior-predictive mixture bet (prior
variance 0.03); the predictive update is the rule \eqref{eq:mp} on the
simulation likelihood, averaged over predictive-resampling paths.}
\label{tab:blr}
\begin{tabular}{llrrr}
\toprule
task & score & false alarms & power & median stop \\
\midrule
drift & refitted point estimate & 0.020 & 0.995 & 202 \\
 & BLR, logistic & 0.022 & 0.985 & 214.5 \\
 & BLR, simulation likelihood & 0.018 & 0.958 & 228.5 \\
 & predictive update, simulation likelihood & 0.028 & 0.897 & 234 \\
bandit & refitted point estimate & 0.028 & 1.000 & 49 \\
 & BLR, logistic & 0.030 & 1.000 & 38 \\
 & predictive update, simulation likelihood & 0.013 & 1.000 & 57 \\
\bottomrule
\end{tabular}
\end{table}

\subsection{Comparison with concurrent methods}\label{sec:exp-concurrent}

The concurrent methods use the same bet and differ in how the score is
chosen, so we run them with the same simulations, mixture over $\lambda$ and
data. For \citet{forre2026anytime} we use the score refitted every ten steps by
maximising the penalised empirical growth, and a score fitted in advance on
simulations from a guessed alternative. For \citet{cortinovis2026slice} we use
SLICE with a single null: an amortised neural ratio network trained on
$2 \times 10^5$ simulations from the simulator's parametrised family
(curvature $b$ for the drift, a location shift per arm for the bandit),
evaluated at a predictable estimate of the parameter. In the bandit both are
run on simulations under the chosen arm, and once as published, with the twin
on random arms. Each task also has an alternative outside SLICE's family: in
the drift the first coordinate is 20\% wider, in the bandit the most pulled
arm keeps its mean and has 50\% larger noise.

\begin{table}[t]
\centering
\caption{Comparison with \citet{forre2026anytime} (F\&B) and
\citet{cortinovis2026slice} (SLICE). Entries: power and, in brackets, median
stopping time; ``in family'': the alternatives of
Sections~\ref{sec:exp-drift} and~\ref{sec:exp-bandit}; ``out of family'':
departures that SLICE's family does not contain. False alarms: range over the
two tasks. Time: computing time per replication on one core, drift / bandit,
excluding the training of SLICE's network. All variants are in the
Supplement.}
\label{tab:concurrent}
\small
\begin{adjustbox}{max width=\textwidth}
\begin{tabular}{lrrrrrr}
\toprule
 & \multicolumn{2}{c}{drift} & \multicolumn{2}{c}{bandit} & false & time \\
score & in family & out of family & in family & out of family & alarms & (s) \\
\midrule
ours, BLR logistic & 0.99 (214.5) & 0.73 (321.5) & 1.00 (38) & 1.00 (112) & 2.0--2.2\% & 0.3 / 0.2 \\
ours, simulation likelihood & 0.96 (228.5) & 0.87 (228) & 1.00 (55) & 0.99 (157) & 1.0--1.8\% & 0.3 / 0.15 \\
F\&B, refitted on the past & 0.99 (210) & 0.91 (228) & 1.00 (61) & 0.99 (167) & 1.5--1.8\% & 1.0 / 0.65 \\
F\&B, fitted in advance & 1.00 (57.5) & 0.01 & -- & -- & 4.8\% & -- \\
SLICE, neural ratio, $G = 1$ & 1.00 (66.5) & 0.01 & 1.00 (35) & 0.56 (258) & 1.5--3.3\% & 0.45 / 0.4 \\
F\&B, twin on random arms & -- & -- & \multicolumn{2}{c}{false alarms 100\%} & -- & -- \\
\bottomrule
\end{tabular}
\end{adjustbox}
\end{table}

In Table~\ref{tab:concurrent}, conditioning on the action is necessary for
every score: run as published, the refitted score of
\citet{forre2026anytime} rejects the true null in every bandit run. Scores
that know where the departure lies are fastest inside the family, SLICE after
67 steps on the drift and a score fitted on the true alternative after 58,
against about 210 for scores learned from the data; but outside the family
their power falls to 0.01 on the drift (Remark~\ref{rem:family}), while every
score learned from real against simulated outcomes keeps it. Among the learned
scores, the refit of \citet{forre2026anytime} is slightly more powerful on the
drift task, markedly so against our logistic learner outside the family (0.91
against 0.73), and ours are faster on the bandit (38 against 61 steps inside
the family, 112 against 167 outside). The gap outside the family comes from
the logistic learner, not from the method. On the simulation likelihood, the
objective that the refit maximises, our update has the same median stopping
time as the refit (228 steps), plausibly because the logistic learner's probit
shrinkage keeps its bets small for longer when the signal is weak. Ours cost a
third as much and carry the pathwise guarantee of Theorem~\ref{thm:simlik}.
They average over the uncertainty about the score's parameter, whereas a sequence of refits at a maximiser is in general not the
predictive sequence of any exchangeable model
\citep{fortini2000exchangeability,fortini2024exchangeability}.

\begin{remark}[Inside and outside the family]\label{rem:family}
SLICE bets with $s_t = \log (p_{\hat\theta_{t-1}}/p_{\theta_0})$ within a
parametrised family $\{P_\theta\}$ with null $\theta_0$, so its tilted null is
$P_{\hat\theta_{t-1}}$. Outside the family, $\hat\theta_t$ converges to the
projection $\theta^* = \arg\min_\theta \KL(Q\|P_\theta)$ and the growth tends to
$\KL(Q\|P_{\theta_0}) - \KL(Q\|P_{\theta^*})$ (Theorem~\ref{thm:growth}(ii)).
When $\theta^* = \theta_0$ the score tends to zero and the wealth stops
growing, however large the departure and however good the network. Both
out-of-family alternatives are of this kind: in the drift no curvature $b$
produces a wider first coordinate and the projection is the null itself, and
in the bandit a location shift cannot produce a larger noise. Our features
contain $x_1^2$ and the squared outcome of each arm, so a change of scale moves
the tilted null towards $Q$. Details, and the choice between the NRE-A and
NRE-C estimators \citep{miller2022contrastive}, are in the Supplement.
\end{remark}

\subsection{Auditing a language-model API with adaptively chosen prompts}\label{sec:exp-llm}

Many language models are used through an API, a service that returns the
model's reply to a text, the prompt, without showing which model produced it.
A provider may advertise a model $P$ and answer some queries with a cheaper
substitute, such as $P$ at lower numerical precision or a smaller model. An
audit checks this from the outside, comparing the API's replies with replies
of $P$ to the same prompts, run locally. The API is the real system, the local
copy of $P$ is the simulator and the prompt is the action; neither gives
probabilities, only replies.

As in the routing audit of \citet{cortinovis2026slice}, the prompts ask for
random output (eight random digits, an integer between 1 and 100, a colour,
six coin flips), so that a substitute shows up as a change in the frequencies
of the replies rather than as a wrong answer. The advertised model is
Qwen2.5-1.5B-Instruct \citep{qwen2024qwen25} stored with 8 bits per weight;
the substitutes are the same model with 4 and 3 bits and the smaller
Qwen2.5-0.5B-Instruct. All replies are real outputs of \texttt{llama.cpp}
\citep{gerganov2023llamacpp} at temperature one; only the routing, the choice
of the queries that go to the substitute, is simulated. Either 30\% of all
queries go to the substitute, or 50\% of the colour and number queries and
none of the others. In a last scenario the API answers every query with a
sharpened $P$, which gives $P$'s frequent replies more often, as a lower
temperature would: the most frequent colour goes from 23\% to 44\% of the
replies.

At each step the auditor chooses a prompt, sends it to the API and generates
$M = 8$ replies of $P$ to the same prompt. The prompt is chosen adaptively,
from the bets won so far. For each prompt the auditor computes the average log
growth of its past bets, adds a bonus that shrinks as the prompt is asked more
often, and asks the prompt with the largest sum, so that it tries prompts it
knows little about as well as those that have paid; this is the upper
confidence bound (UCB) rule of bandit problems \citep{lattimore2020bandit},
used here without tuning. No prompt is asked in fewer than a tenth of the
queries: whenever one falls below that share, it is asked next, so that the
score keeps learning on every prompt.

We compare five scores. Two are ours, both on hashed character $n$-grams
with one block of weights per prompt: the logistic BLR of
Section~\ref{sec:blr}, our default, and the simulation-likelihood rule of the
same section, the one covered by Theorem~\ref{thm:simlik}. They reweight the
replies of $P$ prompt by prompt and need no reply of the substitute. The score
of \citet{forre2026anytime} is used in its two versions: refitted every ten
steps on the whole history by maximising the empirical growth, on the same
features, and fitted in advance to separate a guessed substitute $R$ from
$P$. SLICE assumes that $R$ answers a fraction $\theta$ of the queries, the
same for all prompts, and estimates $\theta$. The score fitted in advance and
SLICE learn $R$ before the audit from 500 replies per prompt of the 4-bit
model, so they guess wrong for the 3-bit and 0.5B substitutes; routing that
depends on the prompt lies outside SLICE's family, and the sharpened $P$ lies
outside it in a stronger sense, since it is closer to $P$ than to $R$.

\begin{table}[t]
\centering
\caption{Auditing a language-model API (Section~\ref{sec:exp-llm}): median
number of queries to an alarm (fewest in bold), rounded to an integer, over
the runs that reject within 400 queries, with the power when below one;
``never'': no run rejects. Routing: ``all'', the substitute answers 30\% of
all queries (the sharpened $P$ answers every query); ``two'', it answers 50\%
of the colour and number queries only. SLICE and the score fitted in advance
are given replies of the 4-bit substitute; the refitted score and ours are
given none. False alarms: rejections of the true null within 400 queries
(Monte Carlo standard error about 1.1\%); a twin that answers a random prompt
raises a false alarm in every run. $^\ddagger$Average of the wealths of ours
(logistic) and of the score fitted in advance. $^*$For the exact posterior;
the one-step Gaussian update used here is covered by
Proposition~\ref{prop:approx}.}
\label{tab:llm}
\small
\setlength{\tabcolsep}{4pt}
\begin{adjustbox}{max width=\textwidth}
\begin{tabular}{llrrrrrl}
\toprule
 & & \multicolumn{4}{c}{substitute} & false & growth \\
\cmidrule(lr){3-6}
score & routing & 4-bit & 3-bit & 0.5B & sharpened $P$ & alarms & guarantee \\
\midrule
\multicolumn{8}{l}{\emph{prompts drawn uniformly (as published)}} \\
SLICE & all & 82 & 149 (0.96) & 110 (0.99) & never & 0.0\% & -- \\
 & two & 76 &  & 107 &  & & \\
\addlinespace[3pt]
F\&B, fitted in advance & all & 44 & 72 & 70 & never & 3.5\% & -- \\
 & two & 39 &  & 59 &  & & \\
\addlinespace[3pt]
F\&B, refitted & all & 251 (0.83) & 170 (0.98) & 102 & 93 & 3.8\% & -- \\
 & two & 200 (0.93) &  & 75 &  & & \\
\addlinespace[3pt]
ours, logistic BLR & all & 255 (0.72) & 209 (0.96) & 123 & 86 & 5.3\% & -- \\
 & two & 219 (0.83) &  & 77 &  & & \\
\addlinespace[3pt]
ours, simulation likelihood & all & 254 (0.69) & 195 (0.99) & 100 & 151 & 4.3\% & Thm~\ref{thm:simlik}$^*$ \\
 & two & 219 (0.85) &  & 62 &  & & \\
\midrule
\multicolumn{8}{l}{\emph{prompts chosen by UCB (our design)}} \\
SLICE & all & 72 & 109 (0.97) & 95 & never & 0.0\% & -- \\
 & two & 57 &  & 57 &  & & \\
\addlinespace[3pt]
F\&B, fitted in advance & all & \textbf{41} & \textbf{50} & \textbf{48} & never & 2.5\% & -- \\
 & two & \textbf{30} &  & \textbf{23} &  & & \\
\addlinespace[3pt]
F\&B, refitted & all & 177 (0.98) & 131 (0.98) & 94 & \textbf{78} & 3.3\% & -- \\
 & two & 107 &  & 51 &  & & \\
\addlinespace[3pt]
average$^\ddagger$ & all & 50 & 66 & 56 & 95 & 2.8\% & -- \\
 & two & 39 &  & 29 &  & & \\
\addlinespace[3pt]
ours, logistic BLR & all & 218 (0.96) & 162 (0.99) & 109 & \textbf{78} & 3.0\% & -- \\
 & two & 136 &  & 52 &  & & \\
\addlinespace[3pt]
ours, simulation likelihood & all & 193 (0.92) & 138 (0.99) & 83 & 130 & 1.5\% & Thm~\ref{thm:simlik}$^*$ \\
 & two & 119 (0.99) &  & 36 &  & & \\
\bottomrule
\end{tabular}
\end{adjustbox}
\end{table}

\begin{table}[t]
\centering
\caption{Cost of reaching the alarm in the audit of
Section~\ref{sec:exp-llm}, with the prompts chosen by UCB: medians over the
runs of Table~\ref{tab:llm} that reject (lowest in bold). Replies: replies
generated by language models, those prepared before the audit (thousands,
column 9) plus $M = 8$ per query. Computing: time spent on the bets and the
updates up to the alarm, on one core. Milliseconds per query: computing time
over 400 and 1600 queries under the null; it grows with the history for the
refit, which revisits the whole history at every refit, and more slowly for
SLICE, which re-estimates its routing fraction from all past replies.
Generating the replies takes 2 to 8 seconds per query on four CPU cores.
$^\ddagger$As in Table~\ref{tab:llm}.}
\label{tab:llmcost}
\small
\setlength{\tabcolsep}{4pt}
\begin{adjustbox}{max width=\textwidth}
\begin{tabular}{llrrrrrrrrr}
\toprule
 & & \multicolumn{2}{c}{4-bit} & 3-bit & \multicolumn{2}{c}{0.5B} & sharpened & prepared & \multicolumn{2}{c}{ms per query} \\
\cmidrule(lr){3-4}\cmidrule(lr){5-5}\cmidrule(lr){6-7}\cmidrule(lr){8-8}\cmidrule(lr){10-11}
score & to the alarm & all & two & all & all & two & $P$, all & ($10^3$) & 400 & 1600 \\
\midrule
SLICE & replies ($10^3$) & 6.6 & 6.5 & 6.9 & 6.8 & 6.5 & never & 6.0 & 1.6 & 2.3 \\
 & computing (s) & 0.12 & 0.09 & 0.17 & 0.15 & 0.09 & never & & & \\
\addlinespace[3pt]
F\&B, fitted in advance & replies ($10^3$) & 6.3 & 6.2 & 6.4 & 6.4 & 6.2 & never & 6.0 & 1.0 & 1.2 \\
 & computing (s) & \textbf{0.05} & \textbf{0.06} & \textbf{0.07} & \textbf{0.06} & \textbf{0.03} & never & & & \\
\addlinespace[3pt]
F\&B, refitted & replies ($10^3$) & \textbf{2.2} & \textbf{1.7} & \textbf{1.8} & 1.5 & 1.2 & \textbf{1.4} & 0.8 & 8.0 & 53.4 \\
 & computing (s) & 0.60 & 0.29 & 0.34 & 0.17 & 0.07 & 0.12 & & & \\
\addlinespace[3pt]
average$^\ddagger$ & replies ($10^3$) & 7.2 & 7.1 & 7.3 & 7.2 & 7.0 & 7.6 & 6.8 & 2.5 & 2.7 \\
 & computing (s) & 0.15 & 0.12 & 0.20 & 0.17 & 0.08 & 0.27 & & & \\
\addlinespace[3pt]
ours, logistic BLR & replies ($10^3$) & 2.5 & 1.9 & 2.1 & 1.7 & 1.2 & \textbf{1.4} & 0.8 & 1.4 & 1.3 \\
 & computing (s) & 0.33 & 0.21 & 0.24 & 0.15 & 0.08 & \textbf{0.11} & & & \\
\addlinespace[3pt]
ours, simulation likelihood & replies ($10^3$) & 2.3 & 1.8 & 1.9 & \textbf{1.5} & \textbf{1.1} & 1.8 & 0.8 & 2.1 & 2.2 \\
 & computing (s) & 0.48 & 0.30 & 0.33 & 0.19 & 0.08 & 0.30 & & & \\
\bottomrule
\end{tabular}
\end{adjustbox}
\end{table}

Choosing the prompts helps every score (Table~\ref{tab:llm}): with our
adaptive design every score needs fewer queries than with uniformly drawn
prompts, in every scenario in which it rejects and up to 61\% fewer, and our
scores gain up to 0.24 in power; the auditor learns to ask for a colour more
often, because that is where the substitutes differ most. Validity requires
the conditioning: comparing the API's reply with replies of $P$ to a random
prompt, rather than to the same prompt, raises a false alarm in every run,
whereas the conditioned monitors raise false alarms in at most 5.3\% of the
runs, within Monte Carlo error of the nominal 5\%.

Knowing the substitute pays when the guess is right and fails when it is not.
With our design the score fitted in advance is the fastest in the five
routing scenarios, after 23 to 50 queries against 36 to 218 for the scores
learned during the audit. When the API sharpens $P$, however, a score fitted
to $R$ has a tilted null close to $R$, hence a growth rate close to
$\KL(Q\|P) - \KL(Q\|R) < 0$ by Theorem~\ref{thm:growth}: it never rejects,
nor does SLICE, which estimates a routing fraction of zero and stops betting.
The three scores learned during the audit reject in every run, after 78 to
130 queries. The average of the wealths of the logistic score and of the
score fitted in advance keeps the best of both: it is within 16 queries of
the faster component in the routing scenarios, and rejects after 95 queries
when the guess is wrong.

Our simulation-likelihood rule and the refit of \citet{forre2026anytime} both
learn the score during the audit and need no guess about the substitute. With
our design the refit needs 5 to 10\% fewer queries on the 4-bit and 3-bit
substitutes and 40\% fewer on the sharpened $P$ (78 against 130), whereas our
rule needs 12 and 29\% fewer on the 0.5B substitute (83 against 94, and 36
against 51); our logistic rule ties with the refit on the sharpened $P$. In
exchange our rule carries the guarantee of Theorem~\ref{thm:simlik}, through
Proposition~\ref{prop:approx} for the one-step update, and its computing time
per query does not grow with the history (Table~\ref{tab:llmcost}): 2.1 and
2.2~ms over 400 and 1600 queries, against 8.0 and 53~ms for the refit, which
revisits the whole history at every refit.

Almost all the running time of the audit goes into the simulator: generating
the $M = 8$ replies of $P$ takes 2 to 8 seconds per query on four CPU cores,
whereas the bet and the update take at most 8~ms over 400 queries, less than
0.5\% of the running time. In these short audits the computing time to the
alarm is therefore at most 0.6~s for every score, and the cost of an alarm is
counted in generated replies, those prepared before the audit plus eight per
query. SLICE and the score fitted in advance prepare 6000 replies, more than
a whole audit, and reach the alarm after 6200 to 6900 replies; the learned
scores prepare 800 replies of $P$, only to standardise the features, and
reach it after 1100 to 2500 (Table~\ref{tab:llmcost}). If one preparation
serves several audits, its cost is shared among them; the growth of the
refit's cost matters instead for long audits and for cheap simulators.

\section{A multi-arm multi-stage trial designed in silico}\label{sec:trial}

\subsection{The trial and its twin}\label{sec:application}

Insulin-dosing algorithms for type~1 diabetes are designed and tested on the
UVA/Padova simulator \citep{dallaman2007meal,kovatchev2009insilico,dallaman2014uva},
accepted by the US Food and Drug Administration for preclinical testing, and
the same simulator is used to design trials of dosing strategies. We use the
open-source \texttt{simglucose} \citep{xie2018simglucose}, a 13-dimensional
model of meal absorption, glucose kinetics and insulin absorption, with three
sources of randomness: insulin sensitivity fluctuating every five minutes,
variable carbohydrate absorption, and the CGM error of
\citet{breton2008analysis}. The state is hidden and the perturbations enter
the dynamics nonlinearly, so the law of a day's CGM trace has no tractable
density.

The trial compares five closed-loop strategies, a control (A) and four
variants of its bolus, correction target and suspension threshold:
aggressive (B), conservative (C), intensified (D) and tight correction (E);
details are in Appendix~\ref{app:details}. The outcome is the time in range
(70--180~mg/dL) over a day with three meals. By the simulator, B and C are
worse than control (time in range 0.79 and 0.84 against 0.91) and D and E
equivalent to it. In each of three stages, 50 new participants, each one of
ten virtual adults, are allocated in permuted blocks among the active arms,
and every enrolled participant is observed for one more day, carrying their
physiological state from one day to the next. After stages 1 and 2 each
experimental arm is dropped for futility when its $z$-statistic against
control falls below $-1$ and $0$ respectively, and its participants are
re-assigned to the best active arm. The trial has 150 participants and 300
participant-days, and every element of the design is chosen from the data.

The twin re-initialises the simulator at each participant's current state and
runs it $M = 16$ times under the arm assigned for that day. The score is the
logistic BLR of Section~\ref{sec:blr} on six summaries of the day relative to
the twin's noise-free path. We compare it with the plug-in bet; a
\emph{stateless} twin, started from rest under the assigned arm, which ignores
the participant's history; a \emph{pooled} twin, started from rest under a
random arm, the pre-trial prediction; and a $t$-test recomputed after every
participant-day. To say under which arm the twin fails, we also run one
e-process per arm, betting only on that arm's participant-days; whether a day
is spent on arm $k$ is decided before its outcome, so each is a test martingale
by Theorem~\ref{thm:valid}, and e-BH \citep{wang2022false} across the arms,
with each arm stopped at $K/\alpha$, controls the false discovery rate under
the dependence created by the shared interim decisions, as in
\citet{xu2021unified}. Under the null the participants are the simulator. Under
\emph{insulin resistance}, insulin sensitivity is 25\% lower in every arm;
under \emph{defensive snacking}, participants on the intensified arms B and D
eat an unannounced 20~g snack at bedtime, while A, C and E remain exactly as
the twin predicts.

\subsection{Results}

\begin{table}[t]
\centering
\caption{A three-stage MAMS trial checked against its in-silico twin (150
participants, 300 participant-days, 100 replications per scenario). Top:
rejection rate and, in brackets, median number of participant-days to the
first alarm. Middle: share of trials with each arm active at the end, and mean
number of re-assigned participants. Bottom: share of trials in which e-BH
flags each arm, at $\alpha = 0.05$.}
\label{tab:trial}
\begin{adjustbox}{max width=\textwidth}
\begin{tabular}{lrrr}
\toprule
monitor & null & insulin resistance & defensive snacking \\
\midrule
MC e-process, twin at the participant's state & 0.02 & 1.00 (4) & 0.99 (32) \\
plug-in bet & 0.47 & 1.00 (5) & 1.00 (28) \\
stateless twin (ignores the history) & 0.95 & 1.00 (4) & 1.00 (33) \\
pooled twin (ignores allocation and history) & 1.00 & 1.00 (8) & 1.00 (58) \\
$t$-test, recomputed & 0.46 & 1.00 (17) & 1.00 (22) \\
\midrule
\multicolumn{4}{l}{\emph{design: share of trials with the arm active at the end}} \\
B / C / D / E & 0.00 / 0.01 / 0.27 / 0.50 & 1.00 / 0.00 / 1.00 / 0.79 & 0.02 / 0.01 / 0.93 / 0.42 \\
participants re-assigned (mean) & 65 & 14 & 43 \\
\midrule
\multicolumn{4}{l}{\emph{per-arm e-processes and e-BH: share of trials in which each arm is flagged}} \\
A / B / C / D / E & 0 / 0 / 0 / 0 / 0.01 & 1 / 1 / 1 / 1 / 1 & 0 / 0.64 / 0 / 0.99 / 0.01 \\
\bottomrule
\end{tabular}
\end{adjustbox}
\end{table}

Only the monitor that conditions on both the allocation and the history is
valid, with 2\% false alarms (Table~\ref{tab:trial}). The stateless twin raises
a false alarm in 95\% of trials, because after the first day participants no
longer start from rest and after a re-assignment their state reflects a
treatment they no longer receive; the pooled twin does so in every trial, and
the recomputed $t$-test and the plug-in bet in 46\% and 47\%. Insulin
resistance is detected after a median of four participant-days. Defensive
snacking is detected in 99\% of trials after 32, and it changes the design: the
snack protects arm D from night-time lows, so D survives to the end in 93\% of
trials instead of 27\% and receives 41\% of the participant-days instead of
17\%, for a reason its in-silico justification does not contain. e-BH names
the arm, flagging D in 99\% of trials and B in 64\%, with a false discovery
rate of 0.3\%; B is found less often because the design usually drops it after
about ten participant-days. With the simulation-likelihood learner the
conclusions are the same and detection slower (51 participant-days instead of
32); the other learners are valid as well (Supplement, Section~S4).

The twin is re-initialised at the participant's true state, which in a real
trial would have to be estimated from CGM and insulin records; plugging in an
estimated state, or sampling it from an approximate filtering distribution,
breaks exact exchangeability, and the guarantees of
\citet{barber2026montecarlo} for approximately exchangeable samples are the
natural tool. An alarm says that the in-silico justification of the design
has failed, not that an arm is unsafe; it would be reported to the data
monitoring committee. A single-stage version with Thompson sampling and the
monitoring of a single patient are in the Supplement.

\section{Discussion}\label{sec:discussion}\label{sec:related}

The Monte Carlo e-process separates validity, the control of false alarms,
from power, the speed at which a wrong simulator is detected; in a classical
test both depend on the test statistic, and learning the statistic from the
data spoils its calibration. Here validity comes only from how the simulator
is run, through exchangeability with fresh simulations under the action taken
and from the current state, and holds for any score and any policy, including
interim analyses and re-assignments. Power comes only from the score, a
sequence of one-step forecasts. Theorem~\ref{thm:growth} measures it by the
distance between the tilted null and the truth, scaled by $M/(M+1)$.
Theorem~\ref{thm:simlik} shows that exact Bayesian updating on the simulation
likelihood loses at most $d\log t$ against the best score of a class chosen in
hindsight, on every path, so that it grows at that score's rate, and since the
posterior-predictive bets multiply to the Bayesian mixture, the wealth is
updated in one step per observation and never recomputed from the past.
Approximations in the score therefore cost power but never validity.

The tilted null also makes a rejection informative. The monitor rejects when
its forecasts, the tilted nulls of its scores, have predicted reality better
than the simulator; as the posterior concentrates, the tilted null approaches
the reweighting of the simulator closest to reality,
$\arg\min_\theta \KL(Q\|Q_\theta)$, and its log ratio to the simulator,
$\theta^\top\phi - \log Z(\theta)$, shows where reality puts more or less mass
than the simulator: a shift in the mean, an excess of variance, a particular
arm. This reading is exploratory rather than inferential. The e-process tests
the global null that the simulator is right, and correlated features can
forecast the same departure equally well, so a large coefficient suggests, but
does not prove, that the simulator fails on that feature. A claim with error
control about one aspect needs its own e-process: a score that depends on the
outcome only through a statistic $f$ is valid for the narrower null that the
simulator reproduces the law of $f$, and several such e-processes, or one per
arm as in the trial, can be combined by e-BH. For a single patient with
unannounced meals, for instance, three such monitors, on night-time fasting
windows, on windows after declared meals and on daytime windows away from
them, test the twin's basal physiology, its model of meal absorption and the
absence of hidden meals. Combined by e-BH, a rejection by the third monitor
alone points to unannounced meals rather than to an inadequate twin, whereas a
rejection on the fasting windows points to the physiology; a monitor that does
not reject, however, gives no evidence that its part of the twin is right.
The tilted null might also drive the design. Under each action $a$, its
divergence from the simulator, $\KL(\tilde T_{t,a}\|P_{t,a})$, is the growth
the monitor expects from that action if its forecast is right, so choosing the
action that maximises it sends the next observation where the monitor believes
the simulator is most wrong. This uses only the past, so it is valid by
Theorem~\ref{thm:valid}; it is a model-based alternative to the upper
confidence bound of Section~\ref{sec:exp-llm}, which uses the growth actually
realised, and it pays only once the forecast is good.

Monte Carlo tests compare a fixed dataset with simulations from the null
\citep{besag1989generalized}, and \citet{fischer2024sequential} make them
anytime-valid in the number of simulations; here the data themselves arrive
over time, with fresh simulations for each observation. The closest work is
\citet{dombowsky2026besag}, who normalise an arbitrary statistic into an
e-value with exchangeable samples, including MCMC samples, and give the
Besag--Clifford e-process for an i.i.d.\ stream; their results on the
likelihood-ratio score are used, with attribution, in
Section~\ref{sec:power}.

Two concurrent papers use a similar betting strategy
for independent data. \citet{forre2026anytime} show that the
likelihood ratio is the optimal score for every $M$, bound the loss from a
finite $M$, and refit the score online by maximising its empirical growth,
leaving dependent streams open. \citet{cortinovis2026slice} invert the
e-process into confidence sequences for the parameters of a simulator, with
amortised neural ratio scores, prove validity given i.i.d.\ covariates, and
note in an appendix that the construction extends to dependent data by
simulating conditionally on the history. Relative to these papers, ours treats
actions chosen from the data and simulators with state, shows that the
conditioning is necessary and quantifies the failure without it, reads
arbitrary scores through the tilted null with the local efficiency
$M/(M+1)$, and updates the score's parameter by Bayes' rule on the simulation
likelihood, with a pathwise regret bound, where both concurrent papers bet at
a point estimate.

SLICE \citep{cortinovis2026slice} answers a different question: which values
of a simulator's parameter are compatible with the data. When the departure lies inside its parametrised family it is
the fastest method we compared; when it does not, its power depends on where
the departure projects onto the family and vanishes when the projection is
the null itself (Remark~\ref{rem:family}), and it needs an amortised training
on the family, of the order of $10^7$ simulations in
\citet{cortinovis2026slice}. A score learned from real against simulated
outcomes needs neither, and pays for this with a slower start when prior
knowledge of the alternative exists: in the audit, a score fitted in advance
on the right substitute stops after 30 to 41 queries, the learned score after
136 to 218. The natural synthesis is to use a pretrained score as the centre
of the prior of the online update, so that the monitor starts as fast as the
pretrained score and corrects it where the guess is wrong, with the guarantees
of Theorems~\ref{thm:valid} and~\ref{thm:simlik} relative to that prior; the
average of the two wealths in Section~\ref{sec:exp-llm}, within 16 queries of
the faster component, is a cruder version of it. In the other direction,
since SLICE bets at the plug-in forecast $P_{\hat\theta_{t-1}}$, its confidence
sequences are a likelihood-free version of the running-estimate confidence
sequences of universal inference \citep{wasserman2020universal}. Averaging the
bets over a posterior instead keeps validity, and a pathwise regret bound for
such confidence sequences would need a posterior computed for each candidate
null from its own simulation likelihood, with scores log-linear in their
parameter; we leave this to future work.

Once the simulator is run under the action taken, the design may be chosen to gather
evidence where it accumulates fastest: in the audit this makes every score
faster, ours and the concurrent ones alike, by 8\% to 61\% in the number of
queries, while without the conditioning on the action the concurrent methods
lose validity altogether. Other anytime-valid methods handle adaptively
collected data, through randomisation tests and e-processes
\citep{nair2023randomization,zampieri2025sequential}, off-policy inference
\citep{waudbysmith2024anytime} or adaptive audits of AI systems
\citep{zhou2026adaptive}, but none uses a simulator as the null; digital-twin
validation by re-initialising the twin from observed states
\citep{ghasemloo2026subtrace} uses the same conditioning with fixed-sample
tests. In bandit multiple testing, one e-process per arm combined by e-BH
controls the false discovery rate under adaptive sampling
\citep{xu2021unified}; there the hypotheses concern the arms' effects and the
sampling is chosen to make the tests powerful, whereas our per-arm monitors
test the twin under the allocation fixed by the trial. Among sequential tests
that bet with a learned predictor, Model-X tests of conditional independence
\citep{shaer2023modelx} use the same mechanism with $M = 1$, drawing a dummy
covariate from its known law; two-sample tests
\citep{podkopaev2023sequential,shekhar2023nonparametric}, kernelised Stein
discrepancy tests \citep{martinez2025sequential} and conformal test
martingales \citep{vovk2003testing,vovk2021testing,prinster2025watch} address
other nulls: two streams compared with each other, a model whose score
function is available, or exchangeability with no reference model.

Computing a marginal likelihood as a product of one-step predictives is the prequential
form of \citet{dawid1984present}, the Bayes universal code of minimum
description length \citep{rissanen1984universal,grunwald2007minimum} and the
basis of sequential Monte Carlo estimates of the evidence
\citep{kong1994sequential,chopin2002sequential}, and it is related to
cross-validation by \citet{fong2020marginal} (Supplement, Section~S5); taking the predictives as the
primitive is the approach of martingale posteriors
\citep{fong2023martingale}. Prequential posteriors
\citep{sinharoy2026prequential} calibrate generative forecasting models
through a scoring-rule loss estimated from simulations. Their aim is to
calibrate a simulator and ours to test it, and the simulation likelihood is
the instance of their construction whose loss, the logarithm of an exactly
fair bet, makes the posterior mixture a test martingale.

Several limitations remain. The null is a simple conditional null; simulators
with unknown calibration parameters need composite-null constructions, such
as universal inference or reverse information projections
\citep{grunwald2024safe}, adapted to samplers. Exact validity requires the
simulator to be run from the state of the real system, which is known in
silico and absent in the audit and the bandit, but must be inferred for a
physical twin. Sampling the state from an exact filtering distribution keeps
validity exact with respect to the observed history; plugging in a point
estimate does not, and an approximate filter, like an MCMC-based simulator,
makes exchangeability only approximate. A persistent error, with divergence
$\delta$ per step between the law of the real outcome and that of the
simulations, can raise a false alarm, but not much before $\log(1/\alpha)/\delta$
steps, since no score grows faster than $\delta$ per step
(Corollary~\ref{cor:falsealarm}(ii)); it matters for long monitoring horizons.
Since the resulting bias recurs at every step, extending the guarantees of \citet{barber2026montecarlo} for a
single approximately exchangeable test to the sequential setting is an open
problem. Theorem~\ref{thm:simlik} compares the monitor with the best score in
a parametric class, not with the true density ratio, and its $d\log t$ regret
is not known to be optimal. It holds for the exact posterior, whereas the
one-step Gaussian update used in practice is covered only through
Proposition~\ref{prop:approx}; and in our experiments the logistic learner,
which has no regret guarantee, is faster than the simulation-likelihood
learner that has one, by about 6\% in stopping time on the drift task and
about 60\% in the trial, because the latter's bets are too cautious. A
tempered posterior $\pi_t \propto \pi_0 L_t^{\gamma}$ with $\gamma > 1$, as in
the prequential posteriors of \citet{sinharoy2026prequential}, would bet more
and is valid for every $\gamma$, since any posterior computed from the past
gives a fair bet; what is lost is the exact factorisation behind
Theorem~\ref{thm:simlik}, and a regret bound for it is the most useful next
step.
Theorem~\ref{thm:simlik} and Proposition~\ref{prop:approx} also assume bounded
features, whereas our experiments standardise polynomial features without
bounding them and clip only the score; the guarantees apply exactly once the
features are clipped. Finally, the experiments are low-dimensional;
high-dimensional simulators call for learned feature maps and structured
posterior approximations. High-energy physics is a natural target, since
searches for new physics already compare collision data with large samples
from the Standard Model simulator using learned classifiers
\citep{dagnolo2019learning}, and the Monte Carlo e-process would make such
comparisons anytime-valid as data accumulate.

Code, results and figures are in the repository \texttt{evalues4generative};
all runs are seeded.

\section*{Data availability}
No external data are used. The replies of the language models in
Section~\ref{sec:exp-llm} were generated with the code below from the publicly
released Qwen2.5 models and are in the \texttt{data} folder of the code
repository; all other data are simulated by the code below.

\section*{Code availability}
All methods, experiments, tables and figures are implemented in Python and
available at \url{https://github.com/elenabortolato/evalues4generative}. The
file \texttt{REPRODUCIBILITY.md} in that repository lists, for every table
and figure, the script and command that produce it, the output file and an
indication of the running time. Every replication is seeded, so the reported
numbers are reproduced exactly with the stated library versions, up to the
rounding differences across processors documented there.

\section*{Disclosure statement}
The author reports there are no competing interests to declare.

\bibliographystyle{plainnat}
\bibliography{references}

\begin{thebibliography}{71}
\providecommand{\natexlab}[1]{#1}
\providecommand{\url}[1]{\texttt{#1}}
\expandafter\ifx\csname urlstyle\endcsname\relax
  \providecommand{\doi}[1]{doi: #1}\else
  \providecommand{\doi}{doi: \begingroup \urlstyle{rm}\Url}\fi

\bibitem[Azuma(1967)]{azuma1967weighted}
Kazuoki Azuma.
\newblock Weighted sums of certain dependent random variables.
\newblock \emph{T{\^o}hoku Mathematical Journal}, 19\penalty0 (3):\penalty0
  357--367, 1967.
\newblock \doi{10.2748/tmj/1178243286}.

\bibitem[Barber and Ramdas(2026)]{barber2026montecarlo}
Rina~Foygel Barber and Aaditya Ramdas.
\newblock {Monte Carlo} testing: non-asymptotic guarantees without joint
  exchangeability.
\newblock \emph{arXiv preprint arXiv:2607.23010}, 2026.

\bibitem[Berti et~al.(2004)Berti, Pratelli, and Rigo]{berti2004limit}
Patrizia Berti, Luca Pratelli, and Pietro Rigo.
\newblock Limit theorems for a class of identically distributed random
  variables.
\newblock \emph{The Annals of Probability}, 32\penalty0 (3):\penalty0
  2029--2052, 2004.
\newblock \doi{10.1214/009117904000000676}.

\bibitem[Besag and Clifford(1989)]{besag1989generalized}
Julian Besag and Peter Clifford.
\newblock Generalized {Monte Carlo} significance tests.
\newblock \emph{Biometrika}, 76\penalty0 (4):\penalty0 633--642, 1989.
\newblock \doi{10.1093/biomet/76.4.633}.

\bibitem[Bissiri et~al.(2016)Bissiri, Holmes, and Walker]{bissiri2016general}
Pier~Giovanni Bissiri, Chris~C. Holmes, and Stephen~G. Walker.
\newblock A general framework for updating belief distributions.
\newblock \emph{Journal of the Royal Statistical Society Series B: Statistical
  Methodology}, 78\penalty0 (5):\penalty0 1103--1130, 2016.
\newblock \doi{10.1111/rssb.12158}.

\bibitem[Bowden and Trippa(2017)]{bowden2017unbiased}
Jack Bowden and Lorenzo Trippa.
\newblock Unbiased estimation for response adaptive clinical trials.
\newblock \emph{Statistical Methods in Medical Research}, 26\penalty0
  (5):\penalty0 2376--2388, 2017.
\newblock \doi{10.1177/0962280215597716}.

\bibitem[Breton and Kovatchev(2008)]{breton2008analysis}
Marc Breton and Boris Kovatchev.
\newblock Analysis, modeling, and simulation of the accuracy of continuous
  glucose sensors.
\newblock \emph{Journal of Diabetes Science and Technology}, 2\penalty0
  (5):\penalty0 853--862, 2008.
\newblock \doi{10.1177/193229680800200517}.

\bibitem[Chaney et~al.(2018)Chaney, Stewart, and
  Engelhardt]{chaney2018algorithmic}
Allison J.~B. Chaney, Brandon~M. Stewart, and Barbara~E. Engelhardt.
\newblock How algorithmic confounding in recommendation systems increases
  homogeneity and decreases utility.
\newblock In \emph{Proceedings of the 12th ACM Conference on Recommender
  Systems}, pages 224--232, 2018.

\bibitem[Chopin(2002)]{chopin2002sequential}
Nicolas Chopin.
\newblock A sequential particle filter method for static models.
\newblock \emph{Biometrika}, 89\penalty0 (3):\penalty0 539--552, 2002.
\newblock \doi{10.1093/biomet/89.3.539}.

\bibitem[Clarke and Barron(1990)]{clarke1990information}
Bertrand~S. Clarke and Andrew~R. Barron.
\newblock Information-theoretic asymptotics of {Bayes} methods.
\newblock \emph{IEEE Transactions on Information Theory}, 36\penalty0
  (3):\penalty0 453--471, 1990.
\newblock \doi{10.1109/18.54897}.

\bibitem[Cortinovis et~al.(2026)Cortinovis, Battaglia, and
  Caron]{cortinovis2026slice}
Stefano Cortinovis, Laura Battaglia, and Fran{\c{c}}ois Caron.
\newblock {SLICE}: Sequential likelihood-free inference via conformal e-values.
\newblock \emph{arXiv preprint arXiv:2610.13855}, 2026.

\bibitem[D'Agnolo and Wulzer(2019)]{dagnolo2019learning}
Raffaele~Tito D'Agnolo and Andrea Wulzer.
\newblock Learning new physics from a machine.
\newblock \emph{Physical Review D}, 99\penalty0 (1):\penalty0 015014, 2019.
\newblock \doi{10.1103/PhysRevD.99.015014}.

\bibitem[Dalla~Man et~al.(2007)Dalla~Man, Rizza, and Cobelli]{dallaman2007meal}
Chiara Dalla~Man, Robert~A. Rizza, and Claudio Cobelli.
\newblock Meal simulation model of the glucose-insulin system.
\newblock \emph{IEEE Transactions on Biomedical Engineering}, 54\penalty0
  (10):\penalty0 1740--1749, 2007.
\newblock \doi{10.1109/TBME.2007.893506}.

\bibitem[Dalla~Man et~al.(2014)Dalla~Man, Micheletto, Lv, Breton, Kovatchev,
  and Cobelli]{dallaman2014uva}
Chiara Dalla~Man, Francesco Micheletto, Dayu Lv, Marc Breton, Boris Kovatchev,
  and Claudio Cobelli.
\newblock The {UVA/PADOVA} type 1 diabetes simulator: new features.
\newblock \emph{Journal of Diabetes Science and Technology}, 8\penalty0
  (1):\penalty0 26--34, 2014.
\newblock \doi{10.1177/1932296813514502}.

\bibitem[Dawid(1984)]{dawid1984present}
A.~Philip Dawid.
\newblock Present position and potential developments: some personal views.
  {Statistical} theory: the prequential approach.
\newblock \emph{Journal of the Royal Statistical Society. Series A (General)},
  147\penalty0 (2):\penalty0 278--292, 1984.
\newblock \doi{10.2307/2981683}.

\bibitem[Dombowsky et~al.(2026)Dombowsky, Engelhardt, and
  Ramdas]{dombowsky2026besag}
Alexander Dombowsky, Barbara~E. Engelhardt, and Aaditya Ramdas.
\newblock {Besag--Clifford} e-values for unnormalized testing.
\newblock \emph{arXiv preprint arXiv:2603.15845}, 2026.

\bibitem[Dulac-Arnold et~al.(2021)Dulac-Arnold, Levine, Mankowitz, Li,
  Paduraru, Gowal, and Hester]{dulacarnold2021challenges}
Gabriel Dulac-Arnold, Nir Levine, Daniel~J. Mankowitz, Jerry Li, Cosmin
  Paduraru, Sven Gowal, and Todd Hester.
\newblock Challenges of real-world reinforcement learning: Definitions,
  benchmarks and analysis.
\newblock \emph{Machine Learning}, 110:\penalty0 2419--2468, 2021.

\bibitem[Durkan et~al.(2020)Durkan, Murray, and
  Papamakarios]{durkan2020contrastive}
Conor Durkan, Iain Murray, and George Papamakarios.
\newblock On contrastive learning for likelihood-free inference.
\newblock In \emph{Proceedings of the 37th International Conference on Machine
  Learning}, volume 119 of \emph{Proceedings of Machine Learning Research},
  pages 2771--2781, 2020.

\bibitem[Fischer and Ramdas(2025)]{fischer2024sequential}
Lasse Fischer and Aaditya Ramdas.
\newblock Sequential {Monte Carlo} testing by betting.
\newblock \emph{Journal of the Royal Statistical Society Series B: Statistical
  Methodology}, 87\penalty0 (4):\penalty0 1200--1220, 2025.
\newblock \doi{10.1093/jrsssb/qkaf014}.

\bibitem[Fong and Holmes(2020)]{fong2020marginal}
Edwin Fong and Chris~C. Holmes.
\newblock On the marginal likelihood and cross-validation.
\newblock \emph{Biometrika}, 107\penalty0 (2):\penalty0 489--496, 2020.
\newblock \doi{10.1093/biomet/asz077}.

\bibitem[Fong et~al.(2023)Fong, Holmes, and Walker]{fong2023martingale}
Edwin Fong, Chris Holmes, and Stephen~G. Walker.
\newblock Martingale posterior distributions.
\newblock \emph{Journal of the Royal Statistical Society Series B: Statistical
  Methodology}, 85\penalty0 (5):\penalty0 1357--1391, 2023.
\newblock \doi{10.1093/jrsssb/qkad005}.

\bibitem[Forr{\'e} and Brenner(2026)]{forre2026anytime}
Patrick Forr{\'e} and Lydia Brenner.
\newblock Anytime-valid simulation-based hypothesis testing.
\newblock \emph{arXiv preprint arXiv:2610.08210}, 2026.

\bibitem[Fortini and Petrone(2024)]{fortini2024exchangeability}
Sandra Fortini and Sonia Petrone.
\newblock Exchangeability, prediction and predictive modeling in {Bayesian}
  statistics.
\newblock \emph{arXiv preprint arXiv:2402.10126}, 2024.

\bibitem[Fortini et~al.(2000)Fortini, Ladelli, and
  Regazzini]{fortini2000exchangeability}
Sandra Fortini, Lucia Ladelli, and Eugenio Regazzini.
\newblock Exchangeability, predictive distributions and parametric models.
\newblock \emph{Sankhy{\=a}: The Indian Journal of Statistics, Series A},
  62\penalty0 (1):\penalty0 86--109, 2000.

\bibitem[Gerganov et~al.(2023)]{gerganov2023llamacpp}
Georgi Gerganov et~al.
\newblock llama.cpp: {LLM} inference in {C/C++}.
\newblock \url{https://github.com/ggml-org/llama.cpp}, 2023.
\newblock Accessed through llama-cpp-python 0.3.36.

\bibitem[Ghasemloo et~al.(2026)Ghasemloo, Eckman, and
  Li]{ghasemloo2026subtrace}
Mohammadmahdi Ghasemloo, David~J. Eckman, and Yaxian Li.
\newblock Subtrace-conditional validation of simulation models and digital
  twins.
\newblock \emph{arXiv preprint arXiv:2607.17088}, 2026.

\bibitem[Gr{\"u}nwald et~al.(2024)Gr{\"u}nwald, de~Heide, and
  Koolen]{grunwald2024safe}
Peter Gr{\"u}nwald, Rianne de~Heide, and Wouter~M. Koolen.
\newblock Safe testing.
\newblock \emph{Journal of the Royal Statistical Society Series B: Statistical
  Methodology}, 86\penalty0 (5):\penalty0 1091--1128, 2024.
\newblock \doi{10.1093/jrsssb/qkae011}.

\bibitem[Gr{\"u}nwald(2007)]{grunwald2007minimum}
Peter~D. Gr{\"u}nwald.
\newblock \emph{The Minimum Description Length Principle}.
\newblock MIT Press, Cambridge, MA, 2007.

\bibitem[Gutmann and Hyv{\"a}rinen(2012)]{gutmann2012noise}
Michael~U. Gutmann and Aapo Hyv{\"a}rinen.
\newblock Noise-contrastive estimation of unnormalized statistical models, with
  applications to natural image statistics.
\newblock \emph{Journal of Machine Learning Research}, 13\penalty0
  (11):\penalty0 307--361, 2012.

\bibitem[Hoeffding(1963)]{hoeffding1963probability}
Wassily Hoeffding.
\newblock Probability inequalities for sums of bounded random variables.
\newblock \emph{Journal of the American Statistical Association}, 58\penalty0
  (301):\penalty0 13--30, 1963.
\newblock \doi{10.1080/01621459.1963.10500830}.

\bibitem[Ie et~al.(2019)Ie, Hsu, Mladenov, Jain, Narvekar, Wang, Wu, and
  Boutilier]{ie2019recsim}
Eugene Ie, Chih-wei Hsu, Martin Mladenov, Vihan Jain, Sanmit Narvekar, Jing
  Wang, Rui Wu, and Craig Boutilier.
\newblock {RecSim}: A configurable simulation platform for recommender systems.
\newblock \emph{arXiv preprint arXiv:1909.04847}, 2019.

\bibitem[Jiang et~al.(2019)Jiang, Chiappa, Lattimore, Gy{\"o}rgy, and
  Kohli]{jiang2019degenerate}
Ray Jiang, Silvia Chiappa, Tor Lattimore, Andr{\'a}s Gy{\"o}rgy, and Pushmeet
  Kohli.
\newblock Degenerate feedback loops in recommender systems.
\newblock In \emph{Proceedings of the 2019 AAAI/ACM Conference on AI, Ethics,
  and Society}, pages 383--390, 2019.

\bibitem[Jones et~al.(2024)Jones, Chang, and Murphy]{jones2024bong}
Matt Jones, Peter Chang, and Kevin Murphy.
\newblock {Bayesian} online natural gradient ({BONG}).
\newblock In \emph{Advances in Neural Information Processing Systems},
  volume~37, 2024.

\bibitem[Kakade and Ng(2005)]{kakade2005online}
Sham~M. Kakade and Andrew~Y. Ng.
\newblock Online bounds for {B}ayesian algorithms.
\newblock In \emph{Advances in Neural Information Processing Systems},
  volume~17, pages 641--648, 2005.

\bibitem[Khan and Rue(2023)]{khan2023bayesian}
Mohammad~Emtiyaz Khan and H{\aa}vard Rue.
\newblock The {Bayesian} learning rule.
\newblock \emph{Journal of Machine Learning Research}, 24\penalty0
  (281):\penalty0 1--46, 2023.

\bibitem[Kong et~al.(1994)Kong, Liu, and Wong]{kong1994sequential}
Augustine Kong, Jun~S. Liu, and Wing~Hung Wong.
\newblock Sequential imputations and {B}ayesian missing data problems.
\newblock \emph{Journal of the American Statistical Association}, 89\penalty0
  (425):\penalty0 278--288, 1994.
\newblock \doi{10.1080/01621459.1994.10476469}.

\bibitem[Kovatchev et~al.(2009)Kovatchev, Breton, Dalla~Man, and
  Cobelli]{kovatchev2009insilico}
Boris~P. Kovatchev, Marc Breton, Chiara Dalla~Man, and Claudio Cobelli.
\newblock In silico preclinical trials: a proof of concept in closed-loop
  control of type 1 diabetes.
\newblock \emph{Journal of Diabetes Science and Technology}, 3\penalty0
  (1):\penalty0 44--55, 2009.
\newblock \doi{10.1177/193229680900300106}.

\bibitem[Lattimore and Szepesv{\'a}ri(2020)]{lattimore2020bandit}
Tor Lattimore and Csaba Szepesv{\'a}ri.
\newblock \emph{Bandit Algorithms}.
\newblock Cambridge University Press, Cambridge, 2020.
\newblock \doi{10.1017/9781108571401}.

\bibitem[Ma and Collins(2018)]{ma2018noise}
Zhuang Ma and Michael Collins.
\newblock Noise contrastive estimation and negative sampling for conditional
  models: consistency and statistical efficiency.
\newblock In \emph{Proceedings of the 2018 Conference on Empirical Methods in
  Natural Language Processing}, pages 3698--3707, Brussels, Belgium, 2018.
\newblock \doi{10.18653/v1/D18-1405}.

\bibitem[Magirr et~al.(2012)Magirr, Jaki, and Whitehead]{magirr2012generalized}
Dominic Magirr, Thomas Jaki, and John Whitehead.
\newblock A generalized {D}unnett test for multi-arm multi-stage clinical
  studies with treatment selection.
\newblock \emph{Biometrika}, 99\penalty0 (2):\penalty0 494--501, 2012.
\newblock \doi{10.1093/biomet/ass002}.

\bibitem[Martinez-Taboada and Ramdas(2025)]{martinez2025sequential}
Diego Martinez-Taboada and Aaditya Ramdas.
\newblock Sequential kernelized {Stein} discrepancy.
\newblock In \emph{Proceedings of the 28th International Conference on
  Artificial Intelligence and Statistics}, volume 258 of \emph{Proceedings of
  Machine Learning Research}, pages 1288--1296, 2025.

\bibitem[McFadden(1974)]{mcfadden1974conditional}
Daniel McFadden.
\newblock Conditional logit analysis of qualitative choice behavior.
\newblock In Paul Zarembka, editor, \emph{Frontiers in Econometrics}, pages
  105--142. Academic Press, New York, 1974.

\bibitem[Miller et~al.(2022)Miller, Weniger, and
  Forr{\'e}]{miller2022contrastive}
Benjamin~Kurt Miller, Christoph Weniger, and Patrick Forr{\'e}.
\newblock Contrastive neural ratio estimation.
\newblock In \emph{Advances in Neural Information Processing Systems},
  volume~35, pages 3262--3278, 2022.

\bibitem[Nair and Janson(2023)]{nair2023randomization}
Yash Nair and Lucas Janson.
\newblock Randomization tests for adaptively collected data.
\newblock \emph{arXiv preprint arXiv:2301.05365}, 2023.

\bibitem[Perdomo et~al.(2020)Perdomo, Zrnic, Mendler-D{\"u}nner, and
  Hardt]{perdomo2020performative}
Juan Perdomo, Tijana Zrnic, Celestine Mendler-D{\"u}nner, and Moritz Hardt.
\newblock Performative prediction.
\newblock In \emph{Proceedings of the 37th International Conference on Machine
  Learning}, volume 119 of \emph{Proceedings of Machine Learning Research},
  pages 7599--7609, 2020.

\bibitem[Podkopaev and Ramdas(2023)]{podkopaev2023sequential}
Aleksandr Podkopaev and Aaditya Ramdas.
\newblock Sequential predictive two-sample and independence testing.
\newblock In \emph{Advances in Neural Information Processing Systems},
  volume~36, pages 53275--53307, 2023.

\bibitem[Prinster et~al.(2025)Prinster, Han, Liu, and Saria]{prinster2025watch}
Drew Prinster, Xing Han, Anqi Liu, and Suchi Saria.
\newblock {WATCH}: Adaptive monitoring for {AI} deployments via
  weighted-conformal martingales.
\newblock In \emph{Proceedings of the 42nd International Conference on Machine
  Learning}, volume 267 of \emph{Proceedings of Machine Learning Research},
  pages 49830--49859, 2025.

\bibitem[{Qwen Team}(2024)]{qwen2024qwen25}
{Qwen Team}.
\newblock Qwen2.5 technical report.
\newblock \emph{arXiv preprint arXiv:2412.15115}, 2024.

\bibitem[Ramdas et~al.(2023)Ramdas, Gr{\"u}nwald, Vovk, and
  Shafer]{ramdas2023game}
Aaditya Ramdas, Peter Gr{\"u}nwald, Vladimir Vovk, and Glenn Shafer.
\newblock Game-theoretic statistics and safe anytime-valid inference.
\newblock \emph{Statistical Science}, 38\penalty0 (4):\penalty0 576--601, 2023.
\newblock \doi{10.1214/23-STS894}.

\bibitem[Rissanen(1984)]{rissanen1984universal}
Jorma Rissanen.
\newblock Universal coding, information, prediction, and estimation.
\newblock \emph{IEEE Transactions on Information Theory}, 30\penalty0
  (4):\penalty0 629--636, 1984.
\newblock \doi{10.1109/TIT.1984.1056936}.

\bibitem[Shaer et~al.(2023)Shaer, Maman, and Romano]{shaer2023modelx}
Shalev Shaer, Gal Maman, and Yaniv Romano.
\newblock Model-{X} sequential testing for conditional independence via testing
  by betting.
\newblock In \emph{Proceedings of the 26th International Conference on
  Artificial Intelligence and Statistics}, volume 206 of \emph{Proceedings of
  Machine Learning Research}, pages 2054--2086, Valencia, Spain, 2023. PMLR.

\bibitem[Shafer(2021)]{shafer2021testing}
Glenn Shafer.
\newblock Testing by betting: a strategy for statistical and scientific
  communication.
\newblock \emph{Journal of the Royal Statistical Society Series A: Statistics
  in Society}, 184\penalty0 (2):\penalty0 407--431, 2021.
\newblock \doi{10.1111/rssa.12647}.

\bibitem[Shekhar and Ramdas(2024)]{shekhar2023nonparametric}
Shubhanshu Shekhar and Aaditya Ramdas.
\newblock Nonparametric two-sample testing by betting.
\newblock \emph{IEEE Transactions on Information Theory}, 70\penalty0
  (2):\penalty0 1178--1203, 2024.
\newblock \doi{10.1109/TIT.2023.3305867}.

\bibitem[Shi et~al.(2019)Shi, Yu, Da, Chen, and Zeng]{shi2019virtual}
Jing-Cheng Shi, Yang Yu, Qing Da, Shi-Yong Chen, and An-Xiang Zeng.
\newblock Virtual-{Taobao}: Virtualizing real-world online retail environment
  for reinforcement learning.
\newblock In \emph{Proceedings of the AAAI Conference on Artificial
  Intelligence}, volume~33, pages 4902--4909, 2019.

\bibitem[Sinha~Roy et~al.(2026)Sinha~Roy, Everitt, Robert, and
  Dutta]{sinharoy2026prequential}
Shreya Sinha~Roy, Richard~G. Everitt, Christian~P. Robert, and Ritabrata Dutta.
\newblock Prequential posteriors.
\newblock \emph{Japanese Journal of Statistics and Data Science}, 2026.
\newblock \doi{10.1007/s42081-026-00374-w}.
\newblock arXiv:2511.17721.

\bibitem[Thomas et~al.(2022)Thomas, Dutta, Corander, Kaski, and
  Gutmann]{thomas2022lfire}
Owen Thomas, Ritabrata Dutta, Jukka Corander, Samuel Kaski, and Michael~U.
  Gutmann.
\newblock Likelihood-free inference by ratio estimation.
\newblock \emph{Bayesian Analysis}, 17\penalty0 (1):\penalty0 1--31, 2022.
\newblock \doi{10.1214/20-BA1238}.

\bibitem[Trippa et~al.(2012)Trippa, Lee, Wen, Batchelor, Cloughesy, Parmigiani,
  and Alexander]{trippa2012bayesian}
Lorenzo Trippa, Eudocia~Q. Lee, Patrick~Y. Wen, Tracy~T. Batchelor, Timothy
  Cloughesy, Giovanni Parmigiani, and Brian~M. Alexander.
\newblock Bayesian adaptive randomized trial design for patients with recurrent
  glioblastoma.
\newblock \emph{Journal of Clinical Oncology}, 30\penalty0 (26):\penalty0
  3258--3263, 2012.
\newblock \doi{10.1200/JCO.2011.39.8420}.

\bibitem[{van den Oord} et~al.(2018){van den Oord}, Li, and
  Vinyals]{oord2018representation}
Aaron {van den Oord}, Yazhe Li, and Oriol Vinyals.
\newblock Representation learning with contrastive predictive coding.
\newblock \emph{arXiv preprint arXiv:1807.03748}, 2018.

\bibitem[Villar et~al.(2015)Villar, Bowden, and Wason]{villar2015multi}
Sof{\'i}a~S. Villar, Jack Bowden, and James Wason.
\newblock Multi-armed bandit models for the optimal design of clinical trials:
  benefits and challenges.
\newblock \emph{Statistical Science}, 30\penalty0 (2):\penalty0 199--215, 2015.
\newblock \doi{10.1214/14-STS504}.

\bibitem[Vovk(2021)]{vovk2021testing}
Vladimir Vovk.
\newblock Testing randomness online.
\newblock \emph{Statistical Science}, 36\penalty0 (4):\penalty0 595--611, 2021.
\newblock \doi{10.1214/20-STS817}.

\bibitem[Vovk and Wang(2021)]{vovk2021evalues}
Vladimir Vovk and Ruodu Wang.
\newblock E-values: calibration, combination and applications.
\newblock \emph{The Annals of Statistics}, 49\penalty0 (3):\penalty0
  1736--1754, 2021.
\newblock \doi{10.1214/20-AOS2020}.

\bibitem[Vovk et~al.(2003)Vovk, Nouretdinov, and Gammerman]{vovk2003testing}
Vladimir Vovk, Ilia Nouretdinov, and Alex Gammerman.
\newblock Testing exchangeability on-line.
\newblock In \emph{Proceedings of the 20th International Conference on Machine
  Learning}, pages 768--775, 2003.

\bibitem[Wang and Ramdas(2022)]{wang2022false}
Ruodu Wang and Aaditya Ramdas.
\newblock False discovery rate control with e-values.
\newblock \emph{Journal of the Royal Statistical Society Series B: Statistical
  Methodology}, 84\penalty0 (3):\penalty0 822--852, 2022.
\newblock \doi{10.1111/rssb.12489}.

\bibitem[Wasserman et~al.(2020)Wasserman, Ramdas, and
  Balakrishnan]{wasserman2020universal}
Larry Wasserman, Aaditya Ramdas, and Sivaraman Balakrishnan.
\newblock Universal inference.
\newblock \emph{Proceedings of the National Academy of Sciences}, 117\penalty0
  (29):\penalty0 16880--16890, 2020.
\newblock \doi{10.1073/pnas.1922664117}.

\bibitem[Waudby-Smith et~al.(2024)Waudby-Smith, Wu, Ramdas, Karampatziakis, and
  Mineiro]{waudbysmith2024anytime}
Ian Waudby-Smith, Lili Wu, Aaditya Ramdas, Nikos Karampatziakis, and Paul
  Mineiro.
\newblock Anytime-valid off-policy inference for contextual bandits.
\newblock \emph{ACM/IMS Journal of Data Science}, 1\penalty0 (3):\penalty0
  10:1--10:42, 2024.
\newblock \doi{10.1145/3643693}.

\bibitem[Whitehead(1997)]{whitehead1997design}
John Whitehead.
\newblock \emph{The Design and Analysis of Sequential Clinical Trials}.
\newblock Wiley, Chichester, 2nd edition, 1997.

\bibitem[Xie(2018)]{xie2018simglucose}
Jinyu Xie.
\newblock Simglucose: a {Python} implementation of the {UVA/Padova} type 1
  diabetes simulator, 2018.
\newblock Software, version 0.2.11, \url{https://github.com/jxx123/simglucose}.

\bibitem[Xu et~al.(2021)Xu, Wang, and Ramdas]{xu2021unified}
Ziyu Xu, Ruodu Wang, and Aaditya Ramdas.
\newblock A unified framework for bandit multiple testing.
\newblock In \emph{Advances in Neural Information Processing Systems},
  volume~34, pages 16833--16845, 2021.

\bibitem[Zampieri(2025)]{zampieri2025sequential}
Fernando~G. Zampieri.
\newblock Sequential randomization tests using e-values: applications for trial
  monitoring.
\newblock \emph{arXiv preprint arXiv:2512.04366}, 2025.

\bibitem[Zhao et~al.(2020)Zhao, Queralta, and Westerlund]{zhao2020sim}
Wenshuai Zhao, Jorge~Pe{\~n}a Queralta, and Tomi Westerlund.
\newblock Sim-to-real transfer in deep reinforcement learning for robotics: A
  survey.
\newblock In \emph{2020 IEEE Symposium Series on Computational Intelligence
  (SSCI)}, pages 737--744, 2020.

\bibitem[Zhou et~al.(2026)Zhou, Vossler, Sivaraman, Mai, and
  Feng]{zhou2026adaptive}
Siyu Zhou, Patrick Vossler, Venkatesh Sivaraman, Yifan Mai, and Jean Feng.
\newblock Adaptive auditing of {AI} systems with anytime-valid guarantees.
\newblock \emph{arXiv preprint arXiv:2605.07002}, 2026.

\end{thebibliography}

\appendix

\section{Notation}\label{app:notation}

\begin{table}[h]
\centering
\caption{Notation. }
\label{tab:notation}
\begin{adjustbox}{max width=\textwidth}
\begin{tabular}{lll}
\toprule
symbol & meaning & where \\
\midrule
$t$, $\F_t$, $\G_t = \F_{t-1} \vee \sigma(A_t)$ & time; information after step $t$; information when the bet at $t$ is placed & Sec.~\ref{sec:setting} \\
$A_t$, $a$ & action chosen by the design (arm, dose, prompt) & Sec.~\ref{sec:setting} \\
$X_t$ & real outcome at time $t$ & Sec.~\ref{sec:setting} \\
$Y_{t,1}, \ldots, Y_{t,M}$; $M$ & simulator draws under $A_t$ from the current state; their number & \eqref{eq:mc} \\
$U_{t,0}, \ldots, U_{t,M}$ & pooled outcomes $(X_t, Y_{t,1}, \ldots, Y_{t,M})$ & Sec.~\ref{sec:simlik} \\
$H_0$ & the simulator reproduces the conditional law under the actions taken & Sec.~\ref{sec:setting} \\
\midrule
$P_t$, $p_t$ & simulator law and density of the next outcome (the null) & Sec.~\ref{sec:setting} \\
$Q_t$, $q_t$ & true law and density of the next real outcome & Sec.~\ref{sec:power} \\
$\tilde T_t$ & tilted null of the score (its forecast), with density $T_t p_t / Z_t$ & \eqref{eq:tilt} \\
$Q_\theta$, $q_\theta$ & model for the truth: the tilt of $P$ by $\theta^\top\phi$ & Sec.~\ref{sec:exactbayes} \\
$\pi_0$, $\pi_t$ & prior and posterior on $\theta$ & Sec.~\ref{sec:exactbayes} \\
$m_t$; $m(x_{1:t})$ & posterior predictive; marginal likelihood & \eqref{eq:prequential} \\
$\nu_t = \mathcal N(\bar\theta_t, S_t^{-1})$ & Gaussian approximation of the posterior (BLR) & \eqref{eq:blr} \\
$P_a$, $\bar P$ & law under action $a$; action-ignoring simulator law & Prop.~\ref{prop:necessary} \\
\midrule
$s_t$, $T_t = e^{s_t}$ & score, chosen from $\G_t$ & \eqref{eq:mc} \\
$Z_t$; $\widehat Z_t$; $Z(\theta)$ & $\E_{P_t}[T_t]$; its Monte Carlo estimate; $\E_P[e^{\theta^\top\phi}]$ & \eqref{eq:tilt}; Sec.~\ref{sec:exactbayes} \\
$B$ & bound on the score, $|s_t| \le B$ & Cor.~\ref{cor:cumulative} \\
$E_t$; $E^{(M)}$; $E_t^\infty$ & Monte Carlo factor; the same in the one-step analysis; its limit as $M \to \infty$ & \eqref{eq:mc} \\
$E_t(\theta)$; $\bar E_t$; $E_t^{\nu}$ & factor of the score $\theta^\top\phi_t$; posterior mixture; mixture under $\nu_{t-1}$ & Sec.~\ref{sec:simlik} \\
$\lambda$, $\Lambda$; $\alpha$ & betting fraction and its grid; level & \eqref{eq:wealth} \\
$W_t$; $\bar W_t$ & wealth; mixture wealth over $\theta$ & \eqref{eq:wealth}; Thm.~\ref{thm:simlik} \\
\midrule
$g_M = g_M(Q, P, T)$; $g_M(\lambda)$ & growth rate with $M$ simulations; with fraction $\lambda$ & \eqref{eq:growthrate} \\
$g_{M,t}$; $g_\infty$ & conditional growth rate at $t$; limit as $M \to \infty$ & \eqref{eq:growthrate} \\
$\rho$ & $\E_Q[T]/Z$ & Thm.~\ref{thm:growth} \\
$\varepsilon$, $\psi$, $h$, $\tilde h$ & local alternative $dQ/dP = 1 + \varepsilon\psi$; score $T = e^{\varepsilon h}$; $h - \E_P[h]$ & Thm.~\ref{thm:growth} \\
$R_a$ & likelihood ratio $dP_a/d\bar P$ & Prop.~\ref{prop:necessary} \\
$\varpi_a$; $u_b$; $K$ & long-run fraction of steps playing $a$; weights of a randomised twin; number of arms & Cor.~\ref{cor:falsealarm} \\
\midrule
$\theta$, $\phi_t$, $z(x)$, $d$ & parameter; features; standardised features; dimension & Sec.~\ref{sec:simlik} \\
$\Theta$, $D$, $G$; $L_t(\theta)$ & parameter set, half its diameter, feature bound; simulation likelihood & Thm.~\ref{thm:simlik} \\
$\bar\theta_t$, $S_t$; $\Sigma_t$; $\theta_t$ & mean and precision of $\nu_t$; covariance of an approximate posterior; parameter of the predictive-first rule & \eqref{eq:blr}; \eqref{eq:mp} \\
$\bar\eta$, $v$, $\kappa$ & predictive mean and variance of the logit $z^\top\theta$; $(1 + \pi v/8)^{-1/2}$ & Sec.~\ref{sec:blr} \\
$\ell_t$, $\nabla_t$, $J_t$; $I_t$ & per-step loss, its gradient and Fisher information; accumulated information & \eqref{eq:mp} \\
$w_k$ & weights of the outcomes: one for $X_t$, $1/M$ for each $Y_{t,j}$ & Sec.~\ref{sec:algorithm} \\
$\mathcal U_t$ & update rule of the predictive & Sec.~\ref{sec:rules} \\
$N$ & number of posterior nodes or predictive-resampling paths & Sec.~\ref{sec:blr} \\
\midrule
$\KL$, $\chi^2$ & Kullback--Leibler and chi-squared divergences & Sec.~\ref{sec:power} \\
\bottomrule
\end{tabular}
\end{adjustbox}
\end{table}

\section{Proofs}\label{app:proofs}

We work conditionally on $\G_t$, so the score is fixed, and write
$\E_{Q\otimes P^M}$ when $X \sim Q$ and $Y_1, \ldots, Y_M$ are i.i.d.\ $P$,
independent of $X$. Two facts recur: $\E_{P\otimes P^M}[E^{(M)}] = 1$
(Proposition~\ref{prop:exact}), and the change of measure
$\E_{Q\otimes P^M}[f/(dQ/dP)(X)] \le \E_{P\otimes P^M}[f]$ for $f \ge 0$, with
equality when $P \ll Q$.

\paragraph{Proof of Theorem~\ref{thm:growth}.}
Let $V = \sum_{j \le M} T(Y_j)$, $E^{(M)} = (M+1)T(X)/\{T(X)+V\}$ and
$E^\infty = T(X)/Z = (d\tilde T/dP)(X)$, so that $g_\infty = \E_Q[\log E^\infty]$.

(i) By Jensen's inequality and the change of measure,
\[
\begin{aligned}
g_M - \KL(Q\|P) = \E_{Q \otimes P^M}\Big[\log\frac{E^{(M)}}{(dQ/dP)(X)}\Big]
&\le \log \E_{Q \otimes P^M}\Big[\frac{E^{(M)}}{(dQ/dP)(X)}\Big] \\
&\le \log \E_{P \otimes P^M}\big[E^{(M)}\big] = 0 .
\end{aligned}
\]

(ii) $V/M \to Z$ almost surely, so
$E^{(M)} = \frac{M+1}{M}\,T(X)/\{T(X)/M + V/M\} \to E^\infty$. If $|s| \le B$,
$T(X)$ and $\{T(X)+V\}/(M+1)$ lie in $[e^{-B}, e^{B}]$, so
$|\log E^{(M)}| \le 2B$ and dominated convergence gives $g_M \to g_\infty$.
Taking $\E_Q$ in $\log(d\tilde T/dP) = \log(dQ/dP) - \log(dQ/d\tilde T)$ gives
$g_\infty = \KL(Q\|P) - \KL(Q\|\tilde T)$.

(iii) If $g_\infty = -\infty$ there is nothing to prove. Otherwise
$\log E^\infty - \log E^{(M)} = \log[\{T(X) + V\}/\{(M+1)Z\}]$. Jensen over
$Y_{1:M}$, with $\E[V] = MZ$, then over $X \sim Q$, with $\E_Q[T]/Z = \rho$:
\[
g_\infty - g_M \le \E_Q\log\Big(1 + \frac{T(X)/Z - 1}{M+1}\Big)
\le \log\Big(1 + \frac{\rho - 1}{M+1}\Big) \le \frac{\rho-1}{M+1}.
\]
This is Theorem~3 of \citet{dombowsky2026besag}.

(iv) Take $\varepsilon < 1/\|\psi\|_\infty$. Let $U_0 = X$, $U_k = Y_k$, and
$\bar h$, $\overline{h^2}$ the averages of $h(U_k)$, $h(U_k)^2$ over
$k = 0, \ldots, M$. Expanding the log of the average of $e^{\varepsilon h(U_k)}$,
with a remainder $R$, $|R| \le C\varepsilon^3\|h\|_\infty^3$ uniformly in the sample,
\[
\log E^{(M)} = \varepsilon h(X) - \log\frac{\sum_k e^{\varepsilon h(U_k)}}{M+1}
= \varepsilon\big(h(X) - \bar h\big) - \frac{\varepsilon^2}{2}\big(\overline{h^2} - \bar h^2\big) - R .
\]
Under $Q \otimes P^M$: $\E[h(X) - \bar h] = \frac{M}{M+1}(\E_Q h - \E_P h)
= \frac{M}{M+1}\varepsilon\E_P[\tilde h\psi]$, since $dQ/dP = 1+\varepsilon\psi$
and $\E_P\psi = 0$; the empirical variance $\overline{h^2} - \bar h^2 \in [0, \|h\|_\infty^2]$
has mean $\frac{M}{M+1}\E_P[\tilde h^2]$ under $P \otimes P^M$, and
$\frac{M}{M+1}\E_P[\tilde h^2] + O(\varepsilon)$ under $Q \otimes P^M$; and
$\E[R] = O(\varepsilon^3)$. Hence
\[
g_M = \frac{M}{M+1}\,\varepsilon^2\Big(\E_P[\tilde h\psi] - \tfrac12\E_P[\tilde h^2]\Big) + O(\varepsilon^3),
\]
and $\E_P[\psi^2] - \E_P[(\psi - \tilde h)^2] = 2\E_P[\tilde h\psi] - \E_P[\tilde h^2]$.
Finally $\KL(Q\|P) = \E_P[(1 + \varepsilon\psi)\log(1 + \varepsilon\psi)]
= \frac{\varepsilon^2}{2}\E_P[\psi^2] + O(\varepsilon^3)$. Similarly,
$\log(d\tilde T/dP) = \varepsilon\tilde h - \frac{\varepsilon^2}{2}\mathrm{Var}_P(h)
+ O(\varepsilon^3)$, so
$\KL(Q\|\tilde T) = \KL(Q\|P) - \E_Q[\log(d\tilde T/dP)]
= \frac{\varepsilon^2}{2}\E_P[(\psi - \tilde h)^2] + O(\varepsilon^3)$, which
gives the efficiency display after the theorem.

\paragraph{Proof of Proposition~\ref{prop:necessary}.}
Given $\G_t$, $T$ is fixed, $X \sim P_a$ and $Y_1, \ldots, Y_M$ are i.i.d.\
$\bar P$. Let $B = \sup T$, $Z = \E_{\bar P}[T] > 0$ and
$\bar T_M = M^{-1}\sum_j T(Y_j)$.

(i) Dividing by $M$,
\[
E_t = \frac{M+1}{M}\cdot\frac{T(X)}{T(X)/M + \bar T_M} \to \frac{T(X)}{Z}
\quad\text{almost surely},
\]
since $(M+1)/M \to 1$, $T(X)/M \to 0$ and $\bar T_M \to Z > 0$ by the strong law.
For the means, let $G_M = \{\bar T_M \ge Z/2\}$. On $G_M$, the denominator is at
least $\bar T_M \ge Z/2$, $T(X) \le B$ and $(M+1)/M \le 2$, so
$E_t \le 2B/(Z/2) = 4B/Z$. Since $\bar T_M \to Z > Z/2$,
$\mathbf 1_{G_M} \to 1$ almost surely, so $E_t^k\mathbf 1_{G_M} \to (T(X)/Z)^k$
with $0 \le E_t^k\mathbf 1_{G_M} \le (4B/Z)^k$, and dominated convergence gives
$\E[E_t^k\mathbf 1_{G_M}] \to \E_{P_a}[T^k]/Z^k$, $k = 1, 2$. On $G_M^c$, $E_t \le M+1$, since the real outcome's
share of the total score is at most one; and Hoeffding's inequality
\citep{hoeffding1963probability} for the $T(Y_j) \in [0, B]$, with deviation
$Z/2$, gives $P(G_M^c) = P(\bar T_M - Z < -Z/2) \le e^{-MZ^2/(2B^2)}$. Hence
$\E[E_t^k\mathbf 1_{G_M^c}] \le (M+1)^k P(G_M^c) \le (M+1)^k e^{-MZ^2/(2B^2)} \to 0$,
an exponential decay beating a polynomial growth.

(ii) By (i), a score $T$ has limit $\E_{P_a}[T]/\E_{\bar P}[T]$, which is one
for every $T$ if $P_a = \bar P$. If $P_a \ne \bar P$, some $A$ has
$P_a(A) > \bar P(A)$ (if $P_a(A) \le \bar P(A)$ for all $A$, the same for $A^c$
gives $\ge$, hence equality), and $T = 1 + \mathbf 1_A$ has limit
$\{1 + P_a(A)\}/\{1 + \bar P(A)\} > 1$. Let $R_a = dP_a/d\bar P$, so
$\int R_a\, d\bar P = 1$ and, by the change of measure,
$\E_{P_a}[R_a] = \int R_a^2\, d\bar P$. Expanding the square,
\[
\chi^2(P_a\|\bar P) = \int (R_a - 1)^2\, d\bar P = \int R_a^2\, d\bar P - 2 + 1 .
\]
If $R_a$ is bounded and bounded away from zero, $T = R_a$ is admissible in (i)
and has limit $\E_{P_a}[R_a]/\E_{\bar P}[R_a] = 1 + \chi^2$. Otherwise we
truncate, $T_c = \min(R_a, c) + 1/c$, which takes values in $[1/c, c + 1/c]$.
As $c \to \infty$,
$\min(R_a, c) \uparrow R_a$, so by monotone convergence
\[
\E_{\bar P}[T_c] \to \int R_a\, d\bar P = 1,
\qquad
\E_{P_a}[T_c] = \int \min(R_a, c) R_a\, d\bar P + \frac1c \to \int R_a^2\, d\bar P,
\]
and the limit for $T_c$ tends to $1 + \chi^2$.

(iii) Set $\mu = (\rho - 1)/2$. Given $\G_t$, the conditional law of $E_t$
depends only on $P_a$, $\bar P$ and $T$, which do not change, so by (i) there
is $M_0$ such that $\E[E_t \mid \G_t] \ge 1 + \mu$ for every $M \ge M_0$ at
every step at which $a$ is played with $T$. Since $T \in [b, B]$, the
numerator of \eqref{eq:mc} is at most $(M+1)B$ and the denominator at least
$(M+1)b$, so $b/B \le E_t \le B/b$ and
$\E[(E_t - 1)^2 \mid \G_t] \le \sigma^2 = (B/b)^2$; then
$\mu/(4\sigma^2) = (\rho-1)b^2/(8B^2)$. It remains to show that a persistent
drift of this kind makes the wealth grow. Let $t_0$ be the time from which,
on $\Omega_0$, these bounds hold, and let
$\xi_i = \log(1 - \lambda + \lambda E_i)$, the log-gain of step $i$;
since $0 < E_i \le M+1$, it lies in $[\log(1-\lambda), \log(1+\lambda M)]$, of
length $C$ say. For $x \ge -1/2$, Taylor's theorem gives
$\log(1+x) = x - x^2/\{2(1+\zeta)^2\}$ with $1+\zeta \ge 1/2$, so
$\log(1+x) \ge x - 2x^2$. With $x = \lambda(E_i - 1) \ge -\lambda \ge -1/2$, on
$\Omega_0$ and for $i \ge t_0$,
\[
\E[\xi_i \mid \G_i] \ge \lambda\,\E[E_i - 1 \mid \G_i] - 2\lambda^2\,\E[(E_i-1)^2 \mid \G_i]
\ge \lambda\mu - 2\lambda^2\sigma^2 \ge \lambda\mu/2,
\]
the last step by $\lambda \le \mu/(4\sigma^2)$. The fluctuations
$\Delta_i = \xi_i - \E[\xi_i \mid \G_i]$ have $\E[\Delta_i \mid \G_i] = 0$ and are
$\G_{i+1}$-measurable, so they are martingale differences, with
$|\Delta_i| \le C$. Azuma--Hoeffding \citep{azuma1967weighted} gives
$P(|\sum_{i \le t}\Delta_i| \ge \varepsilon t) \le 2e^{-\varepsilon^2 t/(2C^2)}$,
summable in $t$, so by Borel--Cantelli $t^{-1}\sum_{i \le t}\Delta_i \to 0$
almost surely, and on $\Omega_0$, where $t_0$ is finite, so does the sum over
$t_0 \le i \le t$. Finally $W_t$ averages $|\Lambda|$ products, one of which is
$\exp(\sum_{i \le t}\xi_i)$, so $W_t \ge |\Lambda|^{-1}\exp(\sum_{i \le t}\xi_i)$ and
\[
\frac{\log W_t}{t} \ge -\frac{\log|\Lambda|}{t} + \frac1t\sum_{i < t_0} \xi_i
+ \frac1t\sum_{t_0 \le i \le t} \E[\xi_i \mid \G_i] + \frac1t\sum_{t_0 \le i \le t}\Delta_i .
\]
On $\Omega_0$ the first two terms tend to zero, the third has $\liminf$ at
least $\lambda\mu/2 = \lambda(\rho-1)/4$ and the last tends to zero.

\paragraph{Proof of Corollary~\ref{cor:falsealarm}.}
(i) When $a$ is played, $X \sim P_a$, the draws are i.i.d.\ $\bar P$ and the
score is $T_a$, so Theorem~\ref{thm:growth}(ii)--(iii) applies with $Q = P_a$,
$P = \bar P$ ($\rho_a < \infty$ as $T_a$ is bounded). The laws do not change and
the actions are finitely many, with frequencies converging almost surely, so
$t^{-1}\sum_{i \le t}\E[\log E_i \mid \G_i] \to \sum_a \varpi_a g_M(a)$, and
Corollary~\ref{cor:cumulative} applies.

(ii) As in Theorem~\ref{thm:growth}(i), with $R_a = dP_a/d\bar P$ and
$e_\lambda = 1 - \lambda + \lambda E_t$,
\[
\E_{P_a \otimes \bar P^M}[\log e_\lambda] - \KL(P_a\|\bar P)
\le \log \E_{\bar P \otimes \bar P^M}[e_\lambda] = 0,
\]
and $\KL(P_a\|\bar P) \le \log\E_{P_a}[R_a] = \log\{1 + \chi^2(P_a\|\bar P)\}$ by
Jensen. Since $|s| \le B$ keeps $\log e_\lambda$ bounded, the martingale strong
law bounds each of the $|\Lambda|$ products, and hence $W_t$. The
likelihood-ratio score $T_a = R_a$, when it is bounded and bounded away from
zero, has $\tilde T_a = P_a$ and attains the bound with $1 \in \Lambda$.

(iii) $\bar P(A) \ge u_a P_a(A)$ for all $A$, so $R_a \le 1/u_a$, which bounds
$\KL = \E_{P_a}[\log R_a]$ and $1 + \chi^2 = \E_{P_a}[R_a]$.

\paragraph{Proof of Proposition~\ref{prop:bayes}.}
With $q_\theta = dQ_\theta/dP$, $\E_P[q_\theta] = 1$, so by Fubini
$Z = \E_P[T_t] = 1$ and $E_t^\infty = T_t(X_t)$. Since
\[
T_t(x) = \frac{\int q_\theta(x) \prod_{i<t} q_\theta(X_i)\,\pi_0(d\theta)}
{\int \prod_{i<t} q_\theta(X_i)\,\pi_0(d\theta)},
\]
the product $\prod_{i \le t} T_i(X_i)$ telescopes to
$\int \prod_{i \le t} q_\theta(X_i)\,\pi_0(d\theta)$. The expansion is
\citet{clarke1990information}, and $\E_Q[\log q_\theta] = \KL(Q\|P) - \KL(Q\|Q_\theta)$.

\paragraph{Proof of Theorem~\ref{thm:simlik}.}
(i) Since $L_t = L_{t-1}E_t$ and $\pi_{t-1} \propto L_{t-1}\pi_0$,
$\int L_t\, d\pi_0 = \int L_{t-1}\,d\pi_0 \cdot \bar E_t$, and by induction
$\int L_t\,d\pi_0 = \prod_{i \le t}\bar E_i = \bar W_t$. As $\pi_{t-1}$ is
$\G_t$-measurable, Theorem~\ref{thm:valid} and Fubini give
$\E[\bar E_t \mid \G_t] = \int \E[E_t(\theta) \mid \G_t]\,\pi_{t-1}(d\theta) = 1$.

(ii) $F = \log L_t$ is concave: each $\log E_i(\theta)$ is linear minus a
log-sum-exp, with gradient $\phi_i(U_{i,0}) - \sum_k w_{ik}\phi_i(U_{i,k})$, where
$w_{ik} \propto e^{\theta^\top\phi_i(U_{i,k})}$ are softmax weights, of norm at
most $2G$. Let $\theta^*$ maximise $F$ on $\Theta$; then
$F(\theta) \ge F(\theta^*) - 2Gt\|\theta - \theta^*\|$. The set
$S_r = \theta^* + \frac{r}{2D}(\Theta - \theta^*) \subseteq \Theta$, $0 < r \le 2D$,
lies within distance $r$ of $\theta^*$ and has $\pi_0(S_r) = (r/2D)^d$, so
\[
\log \bar W_t \ge \log\int_{S_r} e^{F}\, d\pi_0 \ge F(\theta^*) - 2Gtr + d \log\frac{r}{2D},
\]
and $r = d/(2Gt)$ gives the first bound. For an interior maximiser, the
Hessian of $F$ is minus a sum of covariances of vectors of norm at most $G$, so
$F(\theta) \ge F(\theta^*) - \frac{tG^2}{2}\|\theta-\theta^*\|^2$ on the ball $B_r$
of radius $r$ around $\theta^*$, which lies in $\Theta$ and contains $S_r$, so
$\pi_0(B_r) \ge \pi_0(S_r) = (r/2D)^d$; then
$\log \bar W_t \ge F(\theta^*) - \frac{tG^2 r^2}{2} + d\log\frac{r}{2D}$, and
$r = \sqrt{d/(tG^2)}$ gives the second bound.

(iii) Let $B_\Theta = \max_\Theta\|\theta\|$. For fixed $\theta$, $E_i(\theta)$
lies in $[(M+1)/(1 + M e^{2GB_\Theta}),\, M+1]$, so
\[
t^{-1}\textstyle\sum_{i\le t}\{\log E_i(\theta) - g_{M,i}(\theta)\} \to 0
\]
almost surely, as in the proof of
Proposition~\ref{prop:necessary}(iii). By (ii),
$\log \bar W_t \ge \sum_{i \le t} \log E_i(\theta) - d\log t - O(1)$. In the
i.i.d.\ case a countable dense subset of $\Theta$ and the continuity of $g_M$
give the supremum, and the last inequality is Theorem~\ref{thm:growth}(ii)--(iii)
for the score $e^{\theta^\top\phi}$, which applies because
$|\theta^\top\phi| \le GB_\Theta$ makes the score bounded and
$\rho_\theta \le e^{2GB_\Theta}$.

\section{Experimental details}\label{app:details}

\paragraph{Refitted learner.} Balanced logistic regression on standardised
monomials of the observation (degree 4 or 6; for the bandit, per-arm terms
$1, y, y^2$), with a ridge penalty on all coefficients except the intercept.
It is refitted by eight Newton steps on all real observations and up to the
last 4000 simulator draws, every ten steps from step 20; until the first fit
the score is zero and $E_t = 1$.

\paragraph{Logistic BLR.} The same model, with prior $\mathcal N(0, I)$ on the
standardised coefficients and variance 10 on the intercept. Features are
standardised with 2000 (bandit: 3000) simulator draws from a stream
independent of the data. At each step the real observation has weight one and
each of the $M$ simulator draws weight $1/M$, and expectations under the
current posterior use the probit approximation. Scores are clipped to
$[-30, 30]$.

\paragraph{Simulation-likelihood BLR.} Prior $\mathcal N(0, \tau I)$ on the
standardised coefficients (no intercept, which cancels in $E_t(\theta)$), with
$\tau = 1$ or $\tau = 0.03$; the latter was chosen on 24 alternative runs of the
drift task. The drift results in Tables~\ref{tab:blr} and~\ref{tab:concurrent}
use 400 fresh replications, disjoint from these (seeds $5\cdot10^6+i$ under the
null and $6\cdot10^6+i$ under the alternative); the variants in the Supplement
use the original seeds. Expectations under the Gaussian
posterior use 32 fixed antithetic nodes $m + L\xi_s$ with $LL^\top = S^{-1}$,
where $\xi_1, \ldots, \xi_{16}$ are standard normal vectors drawn once and
$\xi_{17}, \ldots, \xi_{32}$ are their negatives, so that the nodes have mean
exactly $m$; the bet is the average of $E_t(\theta)$ over the nodes.

\paragraph{Language-model audit.} Each reply is mapped to 32 features: its
character unigrams and bigrams hashed into 30 bins (CRC32), with counts divided
by the square root of the number of $n$-grams, the length of the reply divided
by ten, and an indicator of whitespace. The score has one block of these
features per prompt, $d = 4 \times 32 + 1 = 129$ weights with the intercept,
standardised with 200 replies of the twin per prompt, and is the logistic
BLR predictive of Section~\ref{sec:blr}. A third of each pool of replies is
held out for the baselines and the rest forms the stream; under the null the
real reply and the $M = 8$ simulations are drawn with replacement from the same
pool. The adaptive policy queries the four prompts in turn for the first eight
steps; afterwards it queries the least used prompt if one has fewer than a
tenth of the queries, and otherwise the prompt maximising
$\bar g_k + \{\log t / n_k\}^{1/2}$, where $n_k$ is the number of queries of prompt
$k$ and $\bar g_k$ the average of $\log(\frac12 + \frac12 E_i)$ over them, computed
from each method's own bets. For the baselines, $r_k/p_k$ is estimated per
prompt by a logistic regression with ridge penalty one and balanced class
weights, on the same features, from the held-out 500 replies of the 4-bit
substitute and 1000 of $P$, with the log ratio clipped to $[-10, 10]$. SLICE's
routing fraction $\hat\theta_{t-1}$ maximises
$\sum_{i<t}\log\{1 - \theta + \theta\,\hat r(X_i)/\hat p(X_i)\}$ over a grid of 201
values in $[0, 1]$, which keeps the argument of the logarithm positive.
The refit of \citet{forre2026anytime} uses the score $\theta^\top z$ on the
same standardised features, refitted every ten steps from step 20 by five
damped Newton steps on the empirical growth penalised by $\|\theta\|^2/2$, the
variance of our prior. The simulation-likelihood rule has prior
$\mathcal N(0, I)$ on the 128 standardised coefficients (no intercept, which
cancels in $E_t(\theta)$) and 32 antithetic nodes, with the standardised
features clipped to $[-10, 10]$, so that they are bounded as
Theorem~\ref{thm:simlik} assumes: a feature that is constant on the twin's
replies, such as whitespace in the replies to one prompt, is otherwise
standardised to about $10^8$ when a substitute's reply has it. In the
sharpened scenario each reply of $P$ is drawn with probability proportional
to the square of its frequency among $P$'s replies to that prompt. Computing
times are measured on one core, over 20 runs of 400 and 5 runs of 1600
queries under the null; the computing time to the alarm in
Table~\ref{tab:llmcost} is measured on the same runs as Table~\ref{tab:llm}.

\paragraph{MAMS trial.} The twin is \texttt{simglucose}
\citep{xie2018simglucose}, the open-source implementation of the 2008 academic
version of the UVA/Padova simulator, through a vectorised port that reproduces
its trajectories to within $5 \times 10^{-4}$~mg/dL. Insulin sensitivity
fluctuates within every five-minute interval ($V_{mx}$ and $k_{p3}$ are
multiplied by the same factor $e^{0.15\eta}$), $C$ grams of carbohydrate are
absorbed as $Ce^{0.2\eta}$, and the CGM has the Johnson-$S_U$ error of
\citet{breton2008analysis}. The five strategies are: A, control, carbohydrate
bolus, correction to 140~mg/dL above 150, basal suspended below 80; B,
aggressive, bolus $\times 1.5$, correction to 100, suspend below 60; C,
conservative, bolus $\times 0.8$, correction to 160, suspend below 90; D,
intensified, bolus $\times 1.2$, correction to 120, suspend below 70; E, tight
correction, bolus $\times 1.0$, correction to 110, suspend below 80. A day runs
from 06:00 to 06:00 with meals of 50, 70 and 80~g at 07:00, 12:30 and 19:00. At
an interim, each active experimental arm is compared with control on all
participant-days so far, and a dropped arm's participants are re-assigned to
the active experimental arm with the highest interim time in range, or to
control if none is left; new participants are allocated only to active arms.
Participant-days are monitored in a fixed order within each stage. The score
uses six summaries of the day relative to the twin's noise-free path (mean and
standard deviation of the residual, time below 70 and above 180~mg/dL, and the
arm), with interactions between the arm and the residual and time-below-range
features. Each arm's e-value is its wealth at the end of the trial, or when it
first reaches $K/\alpha$ with $K = 5$; stopping locks in the rejection and the
stopped wealth is still an e-value by optional stopping.

\paragraph{Seeds and cost.} Replication $i$ uses seed $i$ under the null and
$10^6 + i$ under the alternative, for every method, so methods are compared on
identical real data. On four cores, the drift experiment takes about one minute
per 400 replications, and the MAMS trial about nine minutes per 100 replications of
one scenario.

\section{Timing relative to concurrent work}\label{app:timeline}

The manuscript was in preparation when \citet{forre2026anytime} (arXiv,
6~October 2026) and \citet{cortinovis2026slice} (arXiv, October 2026)
appeared. The conditioned Monte Carlo e-process, the bandit and multi-arm
multi-stage trial experiments, the simulation likelihood with
Theorem~\ref{thm:simlik}, and the per-arm e-BH procedure are recorded in the
version history of the code repository from 29 and 30 September 2026. Both
preprints develop the method for independent data;
\citet[Appendix~C.1.1]{cortinovis2026slice} note the possibility of
conditioning on the past for dependent data. The treatment of actions chosen
from the data and of simulators with state, the necessity of the conditioning
and the use of adaptive actions as a source of power are developed here. The comparisons with both papers in
Sections~\ref{sec:exp-concurrent} and~\ref{sec:exp-llm}, and the remarks
attributing shared results to them, were added after they appeared.

\end{document}